\documentclass[onecolumn,authoryear]{els-mrw} 

\usepackage{amsmath,amssymb,amsfonts,amsthm,makeidx,graphicx}
\usepackage{txfonts}
\usepackage{helvet}

\begin{document}

\chapter{Tidal Disruption Events}

\author[1,2]{Brenna Mockler}%
\author[3,4,5]{Erica Hammerstein}%

\author[6]{Eric R.~Coughlin}%

\author[7]{Matt Nicholl}%

\address[1]{\orgname{Carnegie Science}, \orgdiv{Carnegie Observatories}, \orgaddress{813 Santa Barbara Street, Pasadena, CA, 91101}}
\address[2]{\orgname{University of California, Davis}, \orgdiv{Department of Physics and Astronomy}, \orgaddress{One Shields Avenue, Davis, CA 95616, USA}}
\address[3]{\orgname{University of Maryland}, \orgdiv{Department of Astronomy}, \orgaddress{4296 Stadium Dr, College Park, MD 20742, USA}}
\address[4]{\orgname{NASA Goddard Space Flight Center}, \orgdiv{Astrophysics Science Division}, \orgaddress{8800 Greenbelt Rd, Greenbelt, MD 20771, USA}}
\address[5]{\orgname{Center for Research and Exploration in Space Science and Technology}, \orgaddress{NASA/GSFC, Greenbelt, MD 20771, USA}}
\address[6]{\orgname{Syracuse University}, \orgdiv{Department of Physics}, \orgaddress{Syracuse, NY 13210, USA}}
\address[7]{\orgname{Queen’s University Belfast}, \orgdiv{Astrophysics Research Centre}, \orgaddress{University Road, Belfast
Northern Ireland, BT7 1NN}}


\maketitle

\begin{glossary}[Glossary]
\term{Accretion disk} Rotationally-supported disk of gas, gravitationally bound to a central object. Viscosity in the disk (though poorly understood) allows material to move inwards and increase the mass of the central object. \\

\term{Active Galactic Nucleus} A supermassive black hole that is actively accreting material, leading to bright emission across the electromagnetic spectrum. \\

\term{Apocenter} The point of farthest approach between two interacting bodies. \\

\term{Circularization radius} The semi-major axis of a circular orbit with the same angular momentum as a very eccentric orbit. Often used as an approximation for the radius at which an accretion disk fed from eccentric gas will form. \\

\term{Direct capture radius} The distance from a SMBH at which a freely orbiting object will be swallowed by the SMBH. For a Schwarzschild (spin-zero) SMBH, this distance is $4GM_{\bullet}/c^2$ for a parabolic orbit, while for a maximally spinning SMBH and a prograde, parabolic orbit, the direct capture radius is $GM_{\bullet}/c^2$.

\term{Eddington limit} The theoretical maximum accretion limit of a compact object. Estimated by assuming spherical symmetry and balancing the gravitational pressure of infalling gas with the radiation pressure from the accretion luminosity on the gas (often calculated for a mass to energy efficiency of 0.1). \\

\term{Green valley galaxies} A sub-population of galaxies, usually defined in terms of optical colors or star formation rate, which have evolved off the star formation main sequence but still possess stellar populations significantly younger than those characteristic of elliptical or early-type galaxies. \\

\term{Pericenter} The point of closest approach between  two interacting bodies.

\term{Photosphere} The surface where the optical depth of an object is of order unity. Photons behind the photosphere (from an observer's perspective) will undergo scattering before they can escape, whereas photons above the photosphere can escape and be detected. \\

\term{Post-starburst galaxy} A galaxy which has undergone a recent starburst, usually within the past Gyr and therefore exhibits low or no active star formation but still possesses younger stellar populations produced in the starburst; sometimes called E+A, K+A, or quiescent Balmer strong galaxies. \\

\term{Quiescent galaxy} A galaxy whose supermassive black hole is not actively accreting material. \\

\term{Sphere of influence} The region of the galaxy where the SMBH dominates the kinematics. Generally defined as where the enclosed mass is equal to two times the mass of the central black hole. \\

\term{Supermassive black hole} A black hole with mass in excess of $\sim 10^5 M_{\odot}$ that resides in the central region of a galaxy.

\term{Tidal disruption event} The destruction of a star by the gravitational field of a supermassive black hole.

\term{Tidal radius} The distance from a supermassive black hole at which a star is destroyed by tides.\\



\end{glossary}

\begin{glossary}[Nomenclature]

\begin{tabular}{@{}lp{34pc}@{}}
TDE & Tidal disruption event \\
AGN & active galactic nucleus \\
$e$ & eccentricity of stellar orbit \\
$l$ & specific angular momentum of stellar orbit \\
$l_{\rm lc} = \sqrt{2GM_{\bullet}r_{\rm t}}$ & Loss cone specific angular momentum \\
$r$ & Distance to black hole \\
$r_{\rm p}$ & Pericenter distance to black hole \\
$r_{\rm apo}$ & Apocenter distance to black hole \\
$R_{\star}$ & Radius of tidally disrupted star \\
$M_{\star}$ & Mass of tidally disrupted star \\
$T_{\star} = R_{\star}^{3/2}/\sqrt{GM_{\star}}$ & Stellar dynamical time \\
$\epsilon_{\star} = GM_{\star}/R_{\star}$ & Specific stellar binding energy \\
SMBH/$M_{\bullet}$ & Supermassive black hole/mass thereof \\
$f_{\rm t} = GM_{\bullet}R_{\star}/r^3$ & Tidal gravitational field \\
$r_t$ & Tidal radius \\
$T_{\rm fb} = T_{\star}\left(M_{\bullet}/M_{\star}\right)^{1/2}$ & Fallback time \\
$\beta = r_{\rm t}/r_{\rm p}$ & Impact parameter (a.k.a.~penetration factor) \\
$\dot{M}$ & Fallback rate/Accretion rate \\
$\Delta \epsilon = \epsilon_{\star}\left(M_{\bullet}/M_{\star}\right)^{1/3}$ & Tidal energy spread \\
$R_{\rm g} = GM_{\bullet}/c^2$ ($R_{\rm sch} = 2GM_{\bullet}/c^2 = 2R_{\rm g}$) & Gravitational radius (Schwarzschild radius) \\
$r_{\rm circ} = 2r_{\rm p}$ & Circularization radius \\
$L_{\rm Edd} = 4\pi GM_{\bullet}c/\kappa_{\rm T}$ & 
Eddington luminosity/limit \\
$\varepsilon$ efficiency of conversion from mass to energy \\
$\kappa_{\rm T}$ & Thomson electron scattering opacity \\
$r_{\rm dc}$ & Direct capture radius

\end{tabular}
\end{glossary}

\begin{abstract}[Abstract]

Stars that orbit too close to a black hole can be ripped apart by strong tides, producing a type of luminous transient event called a ``tidal disruption event" (TDE). Tidal disruption events of stars by supermassive black holes (SMBHs) provide windows into the nuclei of galaxies at size scales that are difficult to observe directly outside our own galactic neighborhood. They provide a unique opportunity to study these supermassive black holes under feeding conditions that change dramatically over $\sim$week-month timescales, and that regularly reach super-Eddington mass inflow rates. Their light curves are dependent on the properties of the disrupting black hole, and can be used to help constrain the lower mass end of the SMBH mass function -- a region of parameter space that is difficult to access with classic dynamical mass measurements.
\end{abstract}

\begin{BoxTypeA}[keypoints]{Key Points}
\section*{Nuclear dynamics, the TDE rate, and host galaxies}
\begin{itemize}
    \item The majority of tidal disruption events come from stars orbiting near the sphere of influence of the black hole. Because of this, the majority of stars are disrupted on very eccentric ($e \rightarrow 1$, energy $\rightarrow 0$) orbits.
    \item Observations show tidal disruption events prefer host galaxies that are more centrally concentrated and have colors that indicate stellar populations formed within the past Gyr, superimposed on older stellar populations. These may be signatures of post-merger galaxies.
\end{itemize}

\section*{The Disruption Process}

\begin{itemize}
\item Stars are disrupted when the tidal gravity of the black hole overwhelms the self-gravity of the star. If a star is fully disrupted on a $\sim$zero-energy orbit, its core is destroyed and approximately half of the mass of the star is bound to the black hole, while the other half is unbound. If a star is only partially disrupted, its core survives the encounter and the star's self-gravity influences the orbital energy of the mass that is bound to the black hole after disruption. 

\item The properties of the star, black hole, and the initial orbit will all affect the fallback rate of mass to the black hole after disruption. For the majority of systems, the mass of the black hole, the density profile and compactness of the star, and whether or not the star is fully disrupted are likely to have the largest observational effects on the mass fallback rate.

\end{itemize} 

\section*{Luminosity production and light curves}

\begin{itemize}
\item TDE flares are powered by the dissipation of the kinetic energy of gas from the disrupted star. For material moving near the black hole's event horizon at an appreciable fraction of the speed of light, this is an incredibly efficient form of luminosity production, and regularly produces luminosities near the black hole's Eddington limit.
\item Most TDE light curves exhibit a rapid rise and a slower decline that can be well modeled by a power law and is comparable in timescales ($\sim$weeks - months) and evolution to the theoretical mass fallback rate. The luminosity powering this flare has been argued to come from shocks and/or a forming accretion disk. At late times ($\gtrsim$year), a growing subset of TDEs show a flattening in luminosity that has been attributed to a viscously spreading disk.
\end{itemize}

\section*{The observed temperatures and spectra of TDEs}
\begin{itemize}
    \item The observed radiation comes from both a compact X-ray source, linked to accretion onto the black hole, and from an extended structure emitting most of the optical light. 
    \item The spectra of TDEs exhibit mainly hydrogen, helium and nitrogen emission lines, whose profiles and ratios are determined by the size of the debris atmosphere and the presence of outflowing gas.
\end{itemize}

\end{BoxTypeA}

\section{Introduction}
\label{sec:intro}
The destruction of a star by the tidal gravitational field of a supermassive black hole (SMBH) is known as a tidal disruption event (TDE). TDEs offer rare\footnote{How rare? continue reading to find out!} glimpses into the inner regions (a.k.a.~nuclei) of distant galaxies and represent signposts of the existence of SMBHs. The underlying theory of TDEs is an active area of research, as is connecting that theory to observations. In this chapter we describe what we have come to understand about TDEs in terms of their host galaxies and rates (Section \ref{sec:host}), the process by which a star is destroyed and the immediate aftermath of destruction (Section \ref{sec:disruption}), and the disc formation, emission processes, and spectral characteristics of observed TDEs (Sections \ref{sec:luminosity_production} and \ref{sec: reprocess-spec}). 

\section{Nuclear dynamics, the TDE rate, and host galaxies} \label{sec:host}

\subsection{Two-body Relaxation and the TDE Rate}
In order for a star to be disrupted, its point of closest approach to the SMBH (the pericenter of its orbit) must fall within a critical distance. For the majority of observed TDEs, this critical distance, termed the ``tidal radius", $r_{\rm t}$, is likely between $\sim10-100$ gravitational radii\footnote{$R_g = 5.91 \times 10^6$ km or $\sim 8.5 R_\odot$ for our own Milky Way's SMBH, Sagittarius A*} from the SMBH ($R_g$, of order the black hole's event horizon). For very large black holes ($\gtrsim 10^8 M_\odot$), the tidal radius will move within the black hole's event horizon, and no visible flare will be produced (see Section~\ref{sec:disruption} for more details on the disruption process). 

The tidal radius is a very small fraction of the sphere of influence radius of the SMBH, meaning that the orbits of most destroyed stars are highly eccentric ($e \gtrsim 0.999$). Exactly how stars are put onto such orbits is still debated, but it is generally thought that two-body relaxation \citep{Bahcall1976, Frank1976, Lightman1977, Cohn1978} -- where the orbit of one star is altered due to a series of weak gravitational interactions with other stars -- is the primary mechanism in the majority of galaxies. 

\begin{figure}[htbp]
   \centering
   \includegraphics[width=0.6\textwidth]{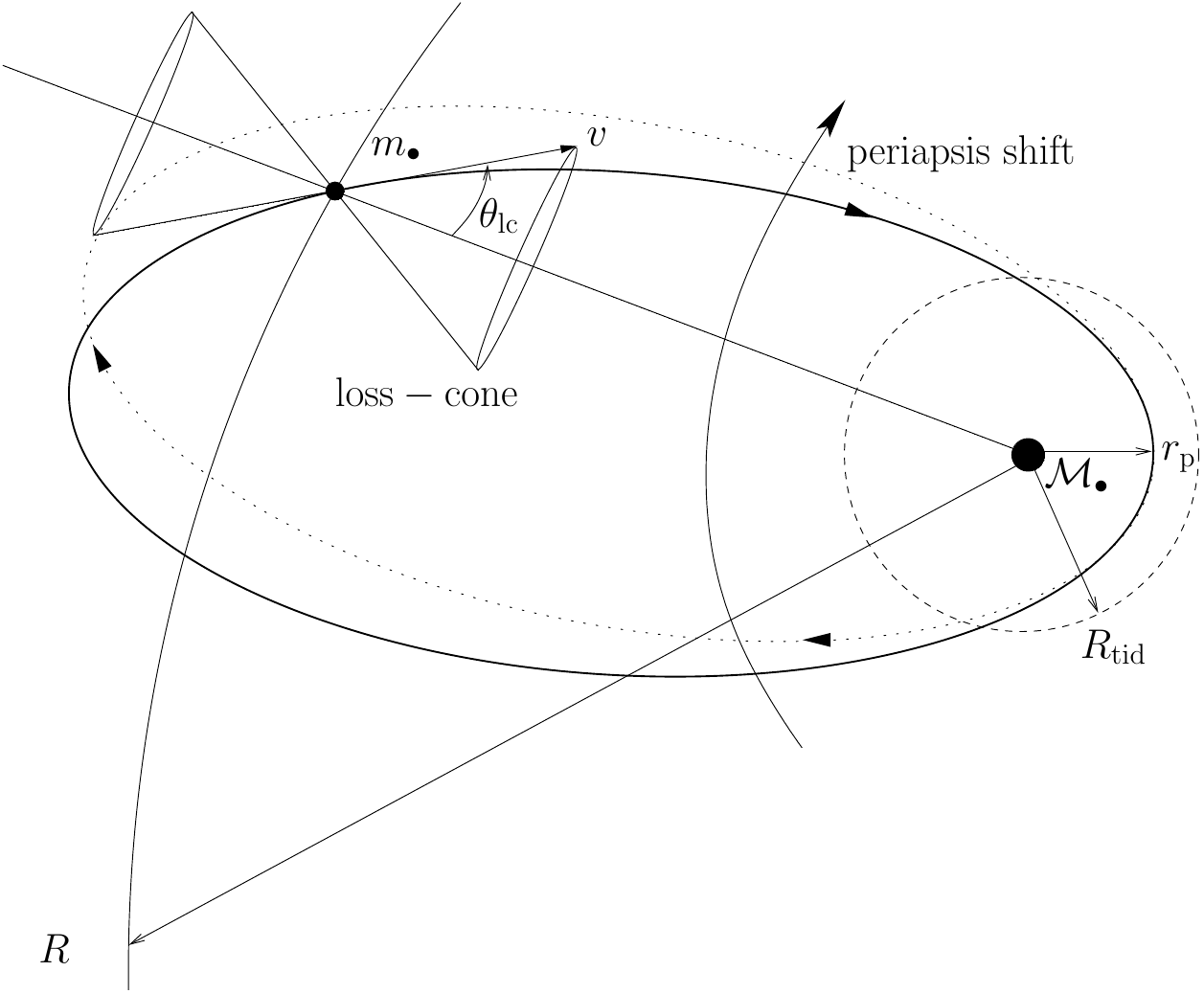} 
   \caption{The loss cone of a star depicted at a point in its orbit around a SMBH. The key takeaway is that the angle of the loss cone $\theta_{\rm lc}$ is defined by the velocity and radial distance vectors of the star. The star will be tidally disrupted if its velocity vector falls into the loss cone. The periapsis shift shows how the orbit will precess under general relativity, which can affect the circularization process of debris post-disruption (discussed later in Section\ref{sec:circularization}), but does not change the loss cone calculation for a given passage. From \citet{Amaro-Seoane2018}.}
   \label{fig:loss-cone}
\end{figure}

We can define the region of parameter space within which stars are destroyed as the black hole's ``loss cone'' (Figure \ref{fig:loss-cone}). This is the collection of orbits for which a star's pericenter radius is less than its tidal radius \citep{Frank1976}. The loss cone is generally defined in angular momentum space. For a star in the supermassive black hole's sphere of influence (the region of the galaxy where the black hole dominates the star's kinematics), its angular momentum can be approximated as follows,
\begin{equation}
    l = \sqrt{GM_\bullet a(1-e^2)},
\end{equation}
where $M_\bullet$ is the mass of the central black hole, $a$ is the semi-major axis, and $e$ is the eccentricity of the orbit. This can be rewritten in terms of the pericenter radius, $r_{\rm p}$, as
\begin{equation}
    l = \sqrt{G M_\bullet r_{\rm p} (1+e)}.
\end{equation}
Because we expect most TDEs to come from stars on highly eccentric orbits ($e \gtrsim 0.999$) we can approximate $e \sim 1$:
\begin{equation}
    l \approx \sqrt{2 G M_\bullet r_{\rm p}}.
\end{equation}

Therefore, a star on a highly eccentric orbit with pericenter distance less than $r_{\rm t}$ has a specific orbital angular momentum less than: 
\begin{equation}\label{eq:losscone}
l_{\rm lc} \leq \sqrt{2GM_{\bullet}r_{\rm t}}.
\end{equation}
Stars on loss cone orbits are disrupted within one orbital period, implying that mechanisms such as two-body relaxation are necessary to ``fill'' the loss cone (i.e., re-populate the region of parameter space with $l \le l_{\rm lc}$) if the rate of TDEs is $\sim$ time independent. 

The effects of two-body relaxation on a star's orbit can be modeled as a random walk in angular momentum space, and therefore the magnitude of the change in angular momentum is proportional to the square root of the number of kicks, or encounters it experiences:
\begin{equation}
     |\Delta l| \propto \sqrt{N_{\rm kicks}}.
\end{equation}
Comparing this to the angular momentum that defines the loss cone ($l_{\rm lc}$) leads to two important regimes::
\begin{enumerate}
    \item $\Delta l > l_{\rm lc}$: The ``pinhole'' regime, in which a star may move into the angular momentum loss cone at some point in its orbit, but is often kicked out by another encounter before it can be disrupted.
    \item $\Delta l < l_{\rm lc}$: The ``diffusive" regime, in which the changes to a star's angular momentum per orbit will be much smaller than the size of the loss cone and it will therefore almost certainly be disrupted before it can experience enough kicks to move it outside of the loss cone again.
\end{enumerate}
The majority of stars that will be disrupted come from the boundary of these two regimes, where $\Delta l \sim l_{\rm lc}$. In this region of parameter space, stars experience enough kicks to quickly re-fill the loss cone, but not so many that they can easily escape after entering but before getting disrupted.\footnote{As it turns out, this critical radius that sources the majority of stars that become TDEs (where $\Delta l \sim l_{\rm lc}$) is of order the radius of the sphere of influence in the SMBH mass range where we find most TDEs ($M_\bullet \sim 10^6 - 10^7 M_\odot$). This is a convenient coincidence, at larger masses this radius will move outside the sphere of influence, and at smaller masses it will move further within the sphere of influence, e.g. \citet{StoneMetzger2016}.} 

The rate at which the loss cone is refilled, and consequently the TDE rate, is related to the distribution function (i.e., the phase space density of specific angular momentum and binding energy) of stars in a galaxy \citep{Magorrian1999} and is therefore sensitive to the stellar density profile in the vicinity of the SMBH and nuclear stellar dynamics \citep{Magorrian1999, Merritt2004, vasiliev_loss-cone_2013, Stone2018}. The rate may also help discriminate between formation scenarios for SMBHs \citep{StoneMetzger2016} or probe the spin distribution of SMBHs \citep{Kesden2012, LawSmith2020}. TDE rates calculated from two-body relaxation typically find a per galaxy rate of $\dot{N}_{\rm tde} \gtrsim10^{-4}$ yr$^{-1}$, with a slight negative dependence on SMBH mass \citep[][for a review]{Wang2004, StoneMetzger2016, Stone2020}.

The first observationally constrained rates -- while not free from selection effects -- were an order of magnitude lower at $\sim 10^{-5}$ yr$^{-1}$ \citep{Donley2002, Esquej2008, Maksym2010, Khabibullin2014, vanVelzen2014, Holoien2016}. Since the advent of wide-field optical surveys, estimates for the observed TDE rate have been more consistent with rates from dynamical theory \citep[e.g.,][]{vanVelzen2018, Hung2018}. This has not necessarily been the case for rates inferred from X-ray selected TDE samples, which have found per-galaxy TDE rates an order of magnitude lower than optical rates \citep{Sazonov2021}. Recent studies have resurrected the rate discrepancy from optical surveys, with \citet{Yao2023} finding a per galaxy rate of $\sim 10^{-5}$ yr$^{-1}$. One possible resolution to this discrepancy is dust obscuration in galactic nuclei (but see \citealt{Roth2021}). \citet{Masterson2024} suggest that the true TDE rate is likely closer to $\sim 6.3 \times 10^{-5}$ yr$^{-1}$, a sum of the X-ray, optical, and infrared rates. \citet{vanVelzen2018} suggest that the flux-limited nature of wide-field surveys leads to many low luminosity TDEs being missed. This may be tested by further characterization of the TDE luminosity function.

In conjunction with the overall rates, the distribution of black holes which produce observable tidal disruption flares is of great interest as it will depend on (and therefore provide us information about) the mass function, occupation fraction and spin distribution of black holes. It will also have some connection to the host galaxies at large as evidenced by scaling relations such as the $M_\bullet - \sigma_\star$ relation. Work to constrain the SMBH masses of TDE hosts using the $M_\bullet - \sigma_\star$ relation \citep[e.g. by][]{Wevers2017, Wevers2019, Hammerstein2023_IFU} has already allowed TDEs to be used to probe the low mass end of the SMBH mass function \citep{vanVelzen2011, Yao2023}. These studies find that the mass distribution of TDE hosts is dominated by SMBHs in the range of $10^6 - 10^7 M_\odot$  \citep[although the $M_\bullet - \sigma_\star$ relation has large uncertainties in this mass range, see][for a review]{Greene2020}. At much larger masses, stars are disrupted inside the SMBH's event horizon (with some dependence on spin, see Section~\ref{sec:highly_destructive}). These ``low-mass" SMBHs are difficult to probe through dynamical mass measurements \citep[e.g.][]{mcconnell_revisiting_2013}, as their spheres of influence are small and the velocities of stars and gas orbiting them are comparatively low. Because of this, most of our black hole mass measurements in this part of parameter space come from active galactic nuclei \citep[AGN, e.g.][]{baldassare_populating_2020}, but we know that most galaxies in our local universe are quiescent \citep[e.g.][]{gair_lisa_2010}. TDEs provide us with an additional avenue towards measuring the occupation fraction and properties of these black holes. 

\subsection{The Host Galaxies of Tidal Disruption Events}

Our ability to study the nuclear regions of a galaxy and the factors that directly affect the TDE rate, such as the nuclear stellar density profile, is limited to only the closest galaxies to us\footnote{The spheres of influence are too small, and there are too many stars in the way to resolve them directly!}. However, the processes that affect these nuclear regions, such as galaxy mergers and star formation, will also affect the large-scale properties of a galaxy that are more readily studied for distant galaxies. These global properties are unlikely to directly affect the TDE rate, as most TDEs are thought to be sourced from stars within the $0.1-10$ pc sphere of influence \citep[][calculated for $10^{6-8} M_\odot$ SMBHs]{StoneMetzger2016}. Nonetheless, the large-scale properties of TDE hosts may correlate with those in the nucleus, in which case studying global properties can still provide insights into the mechanisms that affect the TDE rate.

\begin{figure}[htbp]
   \centering
   \includegraphics[width=0.45\textwidth]{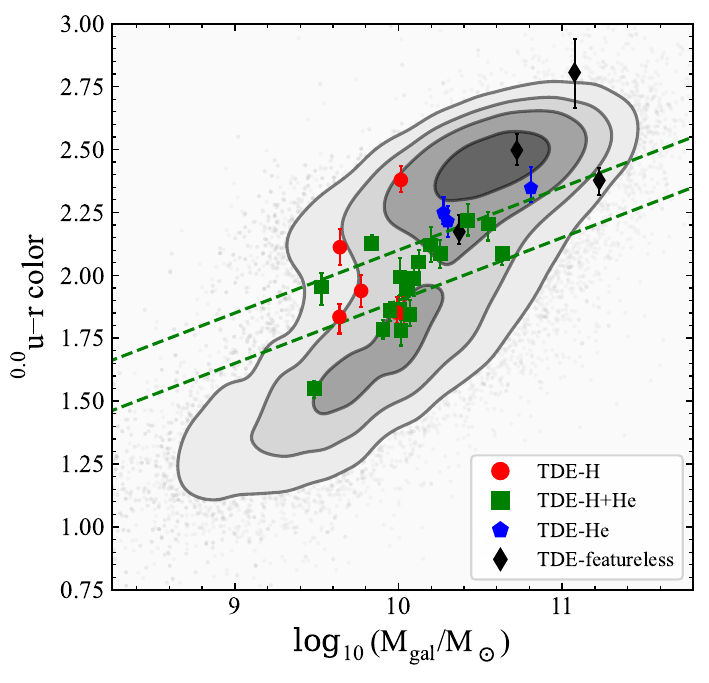} 
   \includegraphics[width=0.45\textwidth]{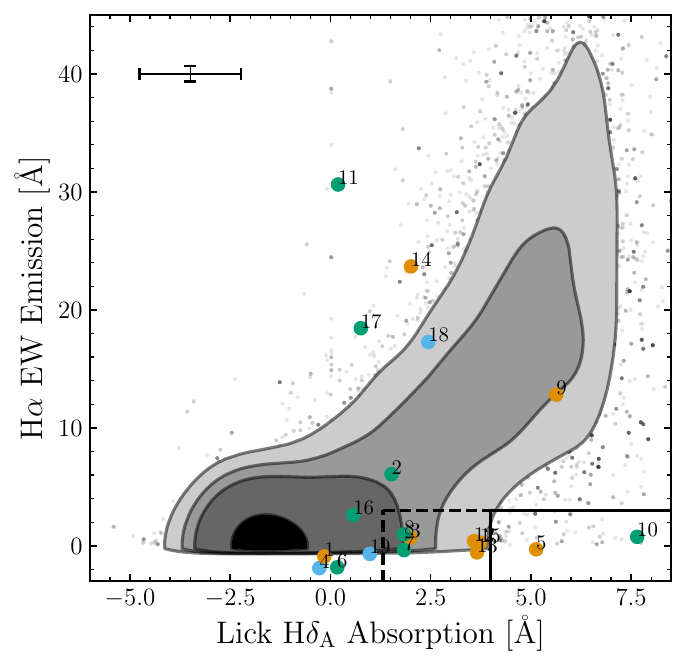} 
   \caption{{\bf Left:} The rest-frame $u-r$ color versus host stellar mass of optically selected TDE host galaxies from the ZTF survey (colored points) compared to a volume-limited sample of SDSS galaxies (background contours). The TDE host galaxies are colored by the spectroscopic sub-type of the TDE (see Section \ref{sec: reprocess-spec} for more). It is often found that $>50\%$ of TDE hosts lie within the two dashed green lines, a region of parameter space known as the green valley. This is in contrast to the much smaller fraction of the general galaxy population ($\sim 13\%$) that resides there. Modified from \citet{Hammerstein2023}.
   {\bf Right:} The Lick H$\delta$ absorption index versus the H$\alpha$ emission equivalent width for a sample of optically selected TDE host galaxies (colored points) compared to a volume-limited background sample of SDSS galaxies (contours). The median uncertainties for the TDE hosts are shown in the top left. The solid boxed region indicates the E+A classification region, whereas the extended dashed region indicates the more loosely defined QBS region. TDE host points are colored by the spectroscopic type of the corresponding TDE. While only two galaxies fall within the E+A region, this corresponds to an E+A overrepresentation factor of $\sim$22 compared to the general galaxy population. This has now been observed in other samples of TDEs discovered in the optical and X-ray. Reproduced with permission from \citet{Hammerstein2021} (their Figure 4).} 
   \label{fig:hostgalaxies} 
\end{figure}

Along these lines, several studies have investigated the large-scale characteristics of TDE host populations. They have found significant trends in properties such as host galaxy stellar mass, optical colors, concentration of the galaxy light profile, and surface brightness. Specifically, when compared to a control sample of galaxies (e.g., a selection from SDSS), TDE hosts have been shown to
possess higher stellar mass surface densities, higher bulge-to-total light ratios, and higher galaxy S\'ersic indices\footnote{The S\'ersic profile \citep{Sersic1963} is defined by $I(R) = I_e \exp \{ -b_n [( R/R_e)^{1/n} - 1 ]\}$, where $R_e$ is the half-light radius, $I_e$ is the intensity at that radius, $n$ is the S\'ersic index, and $b_n$ is a function of the S\'ersic index. Higher S\'ersic indices imply a steeper or more centrally-concentrated light profile, while a lower S\'ersic index implies a shallower light profile. Elliptical galaxies are typically well-described by a S\'ersic profile with $n=4$, otherwise known as the de Vaucouleurs profile \citep{deVaucouleurs1948}, whereas disk or spiral galaxies can be described by a S\'ersic profile with $n=1$, i.e., an exponential profile.} 
than galaxies of similar total stellar masses \citep{Graur2018, LawSmith2017, Hammerstein2021}. This trend in high concentration is not only limited to global properties, but extends down to the inner 30--100 pc in nearby TDE hosts with high-resolution imaging \citet{French2020}. 

TDE hosts have also been associated with specific evolutionary phases in galaxy evolution. TDE hosts selected at both optical and X-ray wavelengths reside in what is known as the ``green valley'' -- between red, passive galaxies (the red sequence) and blue, star-forming spirals (the blue cloud), as shown in Figure~\ref{fig:hostgalaxies} \citep[e.g.,][]{LawSmith2017, Hammerstein2021, Sazonov2021, Yao2023}. This sub-population of galaxies is typically defined in terms of star-formation rate or global galaxy color, as the color is a proxy for the stellar population age and the level of ongoing star formation. Green valley galaxies are expected to lack ongoing star formation as many of the largest, hottest (bluest) stars have died out.

Further analysis of the hosts is possible by looking at their spectra. \citet{Arcavi2014} was the first to note that many of the TDE host spectra are characterized by both a lack of strong emission lines that would indicate active star formation (quantified by the equivalent width of the H$\alpha$ emission), and also the existence of strong Balmer absorption features that are characteristic of young stellar populations, particularly A stars \citep[quantified by the the Lick H$\delta_{\rm A}$ absorption index,][]{Worthey1997}. This indicates that, while there isn’t ongoing star formation, a star formation event occurred within approximately the past Gyr \citep{Dressler1983, Zabludoff1996}. Together, these features are trademarks of a class of galaxies known as E + A (elliptical+A stars), K + A (K stars+A stars), or more generally, post starburst or quiescent Balmer-strong (QBS) galaxies \citep{Dressler1983, Zabludoff1996}.

Post-starburst galaxies are rare -- the exact percentage of galaxies in this class depends on the cuts made on the parameters described above, but it is typically less than 1\% of the general population (see Figure~\ref{fig:hostgalaxies}). Despite this, post-starbursts comprise a relatively large percentage of TDE host populations. For example, \citet{French2016} found that QBS galaxies are overrepresented among TDE hosts by a factor of 33–190 \citep[see also][for other host samples selected from the optical to X-ray]{LawSmith2017, Graur2018, Hammerstein2021, Sazonov2021}.

Interestingly, the preference for both green valley hosts and post-starburst hosts is not seen for TDEs discovered in the infrared \citep{Masterson2024}. Dust extinction has been suggested as playing a role in driving the different host preferences. However, some studies have shown that dust obscuration alone cannot fully account for either the preference for ``green'' TDE hosts or the lack of TDEs observed in low-mass red galaxies \citep{Roth2021}. Further work is needed to understand whether the properties of infrared selected TDEs are similar in other aspects to the optical and X-ray selected TDE host population.

The reason for the host preferences in optical and X-ray selected samples and its connection to increased TDE rates remains uncertain. Attempts have been made to explain it through selection effects in e.g. SMBH mass, optical color, surface brightness, or concentration of the galaxy light profile \citep{LawSmith2017, Hammerstein2021}. For example, work on an early Zwicky Transient Facility (ZTF) sample showed it is possible that the overabundance of post-starburst host galaxies is part of a more general preference for green, centrally concentrated galaxies \citet{Hammerstein2021}.

In addition to explanations related to observational selection effects, a number of physical origins have also been suggested for this enhancement. Post-starburst galaxies and TDE hosts (even those that do not strictly meet the post-starburst classification) are known to have high S\'ersic indices, large bulge-to-total light ratios, and high concentration indices \citep[e.g.,][]{French2020,  Graur2018, Hammerstein2021, LawSmith2017}. These properties are all indicative of high nuclear stellar densities, however, high densities alone are probably not sufficient to explain the host preferences. For example, there is no rate enhancement observed in early-type galaxies, which are also centrally concentrated. Additionally, continuously increasing the density of stars can eventually lead to loss-cone shielding\footnote{This is where the increased rates of strong scatterings (large-angle scatterings, unlike the small-angle scatterings that dominant two-body relaxation) of stars can eject them from the sphere of influence, and in the most extreme case can result in direct collisions that destroy the stars \citep{Teboul2024}. It becomes relevant at power-law indices $\alpha \gtrsim 2$ for density profiles that go as $\rho_* \propto r^{-\alpha}$.}, preventing further TDE rate enhancements. Interestingly, these properties have also been used to suggest that post-starburst galaxies could be the result of a merger-driven starburst \citep{Yang2008}. If a large fraction of TDE hosts experienced recent mergers, this may point to several mechanisms that could work together to boost the rate. Such effects include nuclear stellar over-densities \citep{French2020, StoneMetzger2016, Stone+vV2016}, as well as asymmetries in the stellar potential \citep[e.g.][]{vasiliev_loss-cone_2013, Merritt2004}, and eventually SMBH binaries \citep[e.g.,][]{Ivanov2005, Wegg2011}, all of which can lead to a higher flux of stars into the TDE loss cone (see Section~\ref{sec:openquestions} for more details). It is possible that the TDE enhancement corresponds to a phase in galaxy evolution\footnote{For example, \citet{Hammerstein2023_IFU} find that TDE host galaxies are younger than galaxies with similar kinematics, implying some time-dependence to the nuclear properties.} (of which green, post-starburst galaxies are a part of ), and some combination of the properties described above are together responsible for the observed enhancement.

\section{The Disruption Process}
\label{sec:disruption}

\subsection{Fundamentals and analytic scalings}
\subsubsection{The tidal radius}
\begin{figure}[htbp]
   \centering
   \includegraphics[width=0.6\textwidth]{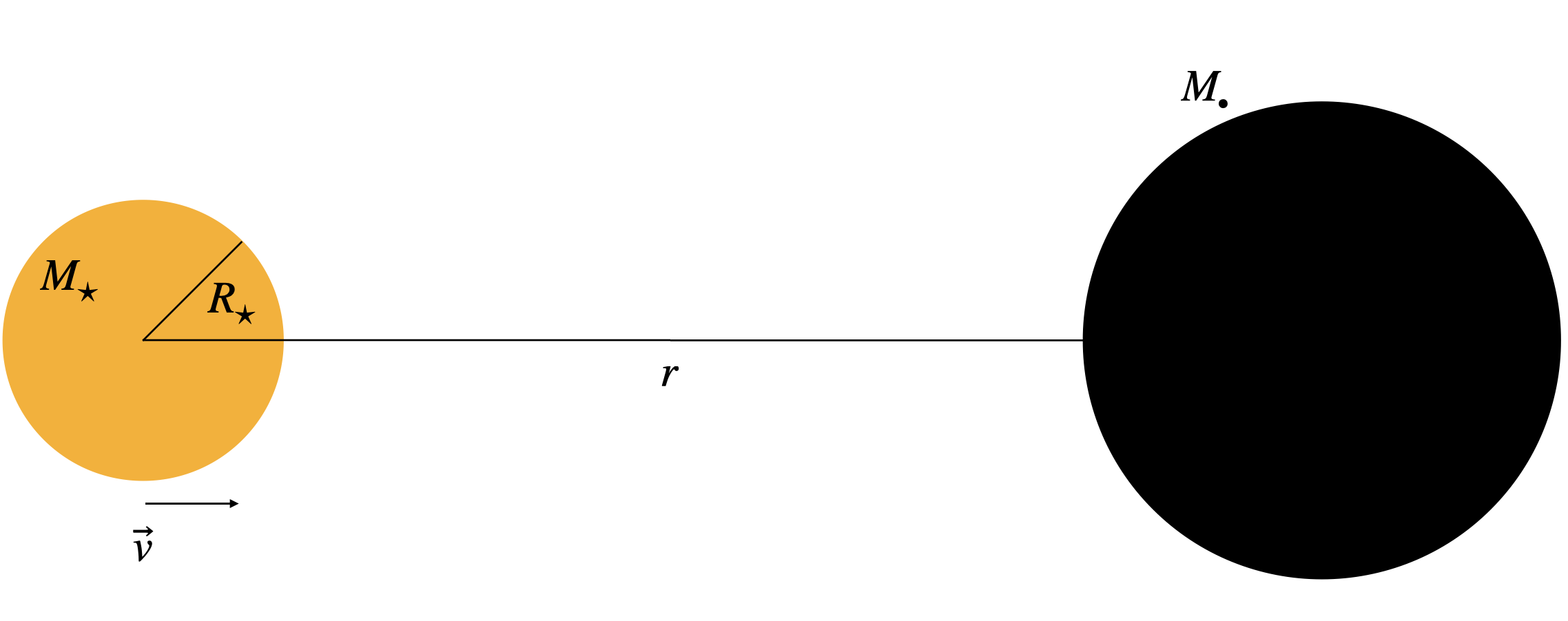} 
   \caption{A star of mass $M_{\star}$ and radius $R_{\star}$ approaches a black hole of mass $M_{\bullet}$; the distance between the center of the star and the black hole is $r$. }
   \label{fig:schema}
\end{figure}

Consider a star of mass $M_{\star}$ and radius $R_{\star}$ approaching a supermassive black hole (SMBH) of mass $M_{\bullet}$, and denote the distance between the center of the star and the SMBH by $r$; see Figure \ref{fig:schema}. The magnitude of the SMBH's gravitational field is $f = GM_{\bullet}/r^2$, and hence the difference in the magnitude of the gravitational field between the near and far sides of the star is
\begin{equation}
\Delta f \equiv f_{\rm t} = \frac{GM_{\bullet}}{\left(r-R_{\star}\right)^2}-\frac{GM_{\bullet}}{\left(r+R_{\star}\right)^2} \simeq \frac{4GM_{\bullet}R_{\star}}{r^3}\left(1+\mathcal{O}\left[R_{\star}/r\right]\right). \label{ftidal}
\end{equation}
The last approximate equality results from Taylor-expanding about the quantity $R_{\star}/r$. This is only accurate when $R_{\star}/r \ll 1$, and is known as the tidal approximation. 
Counterbalancing the tidal field is the self-gravitational field of the star, which near the stellar surface is
\begin{equation}
f_{\rm sg} = \frac{GM_{\star}}{R_{\star}^2}.
\end{equation}
The importance of tides is measured by the ratio of the tidal field to the self-gravitational field:
\begin{equation}
\frac{f_{\rm t}}{f_{\rm sg}} \simeq \frac{M_{\star}}{M_{\bullet}}\left(\frac{R_{\star}}{r}\right)^3. \label{Dfratio}
\end{equation}
If this ratio is $\sim 1$, the star will be significantly tidally distorted. Setting the ratio in Equation \eqref{Dfratio} to 1 and solving for the distance at which this equality is satisfied then yields the canonical definition of the tidal radius:
\begin{equation}
r_{\rm t} = R_{\star}\left(\frac{M_{\bullet}}{M_{\star}}\right)^{1/3}. \label{rtdef}
\end{equation}

For $M_{\bullet} = 10^{6}M_{\odot}$, $r_{\rm t} \gtrsim 100\,R_{\star} \gg R_{\star}$, which is a consistency check on the tidal approximation, i.e., $R_{\star}/r_{\rm t} \ll 1$.

If the point of closest approach of the star to the SMBH is $r_{\rm p}$, it is customary to define the impact parameter $\beta$ as 
\begin{equation}
\beta = \frac{r_{\rm t}}{r_{\rm p}} \label{betadef}.
\end{equation}
Encounters between a star and an SMBH can thus be divided into three regimes: perturbative with $\beta \ll 1$, mildly disruptive with $\beta \sim 1$, and ``deep''/highly destructive with $\beta \gg 1$. In the sections below we discuss numerical investigations of these limits, but we first provide analytic estimates to aid in understanding. 

\subsubsection{The frozen-in approximation}
\label{sec:frozen}
An analytical TDE model, originally due to \citet{lacy82}, is the ``frozen-in'' or ``impulse'' approximation: from Equation \eqref{ftidal}, tidal effects are very weak at distances $r > r_{\rm t}$ and completely dominant for $r < r_{\rm t}$. We therefore assume that the star retains perfect spherical symmetry prior to reaching $r_{\rm t}$, such that each infinitesimal mass element (a.k.a.~fluid element) of which our star is composed moves with the same Keplerian velocity. Upon reaching $r_{\rm t}$, the tidal field is so strong that the stellar self-gravity and pressure can be ignored in comparison to the gravitational field of the SMBH. Thus, after reaching the tidal radius, each fluid element moves ballistically and traces out a Keplerian orbit. The geometry of a Keplerian orbit is established by its specific energy 
\begin{equation}
\epsilon = \frac{1}{2}v^2-\frac{GM}{r}, \label{eps}
\end{equation}
where $v$ and $r$ are the speed and distance of a fluid element, and this energy is conserved, or ``frozen-in,'' at\footnote{The model of \citet{lacy82} and implemented by, e.g., \citet{carter82, bicknell83, rees88}, unambiguously established the distance at which the energy is frozen-in as $r_{\rm t}$, but $r_{\rm t}$ was replaced with $r_{\rm p}$ by, e.g., \citealt{evans89, Ulmer1999, lodato09, strubbe09}. This was then corrected by \citet{stone13, guillochon13}.} $r_{\rm t}$. Each fluid element has speed $\sqrt{2GM/r_{\rm t}}$ at $r_{\rm t}$, but their distances differ from one another by $\sim$ the diameter of the star. Setting $v = \sqrt{2GM/r_{\rm t}}$ and $r = r_{\rm t}\pm R_{\star}$ in Equation \eqref{eps} and making the tidal approximation, the energies of the most-unbound and most-bound debris produced from a TDE are $\pm\Delta\epsilon$, where\footnote{Note that we have already assumed here that $e = 1$ when evaluating the speed at $r_{\rm t}$; if the star is on an orbit with $e \neq 1$, the energy spread is centered around the binding energy of the original orbit.}
\begin{equation}
\Delta \epsilon = \frac{GM_{\bullet}R_{\star}}{r_{\rm t}^2} = \frac{GM_{\star}}{R_{\star}}\left(\frac{M_{\bullet}}{M_{\star}}\right)^{1/3}. \label{epst}
\end{equation}
$GM_{\star}/R_{\star}$ is the energy scale by which the star is bound to itself. Equation \eqref{epst} thus shows that a SMBH imparts energies to the tidally destroyed debris that are orders of magnitude larger than the stellar binding energy.

Fundamental aspects of TDEs that follow directly from Equation \eqref{epst} are:

\begin{enumerate}
\item Half of the material in a TDE is bound to the SMBH, and these fluid elements trace out elliptical orbits with the SMBH at one focus. 
\item The apocenter (point of farthest approach from the SMBH) 
of the most-bound orbit is $\simeq \left(M_{\bullet}/M_{\star}\right)^{1/3}\times r_{\rm t}$, while the pericenter is $\simeq r_{\rm t}$. The orbits of the debris are thus highly elliptical with eccentricities $e \gtrsim 0.99$. 
\item The orbital time of the most-bound debris, a.k.a.~the fallback time, 
is $T_{\rm fb} = 2\pi a^{3/2}/\sqrt{GM_{\bullet}} = \pi R_{\star}^{3/2}/\sqrt{2GM_{\star}}\times \left(M_{\bullet}/M_{\star}\right)^{1/2}$. 
\item Fluid elements with positive specific energy have velocities at infinity of $v_{\infty} = \sqrt{2\epsilon_{+}} = \sqrt{2GM_{\star}/R_{\star}}\times \left(M_{\bullet}/M_{\star}\right)^{1/6}$. 
\end{enumerate}
Of these points, perhaps the most important in the context of observations is the third -- for a $10^6M_{\odot}$ SMBH and a solar-like star, such that $R_{\star} = 1R_{\odot} \sim 7\times 10^{10}$ cm and $M_{\star} = 1M_{\odot} \sim 2\times 10^{33}$ g, the fallback time is $T_{\rm fb} \sim 40$ days. As discussed in Section \ref{fig:emissiontimescale} below, if the material can accrete promptly onto the black hole, this timescale should be reflected in TDE lightcurves.

\subsubsection{The fallback rate}
\label{sec:fallback}
\begin{figure}[htbp] 
   \centering
   \includegraphics[width=0.495\textwidth]{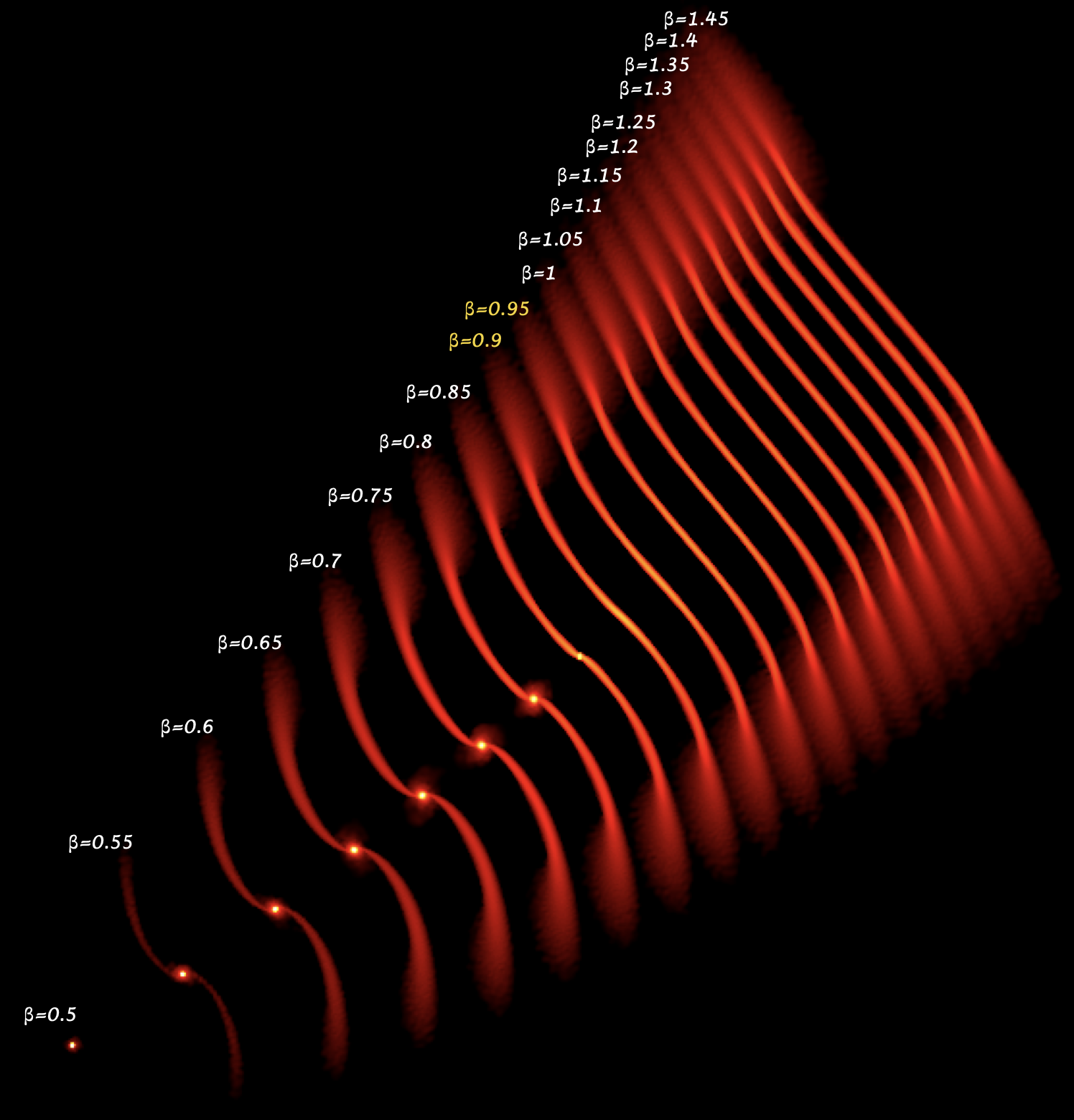} 
   \caption{The debris stream produced from the partial and complete disruption of a $5/3$ polytropic star. Figure adapted from \citet{mainetti17}. }
   \label{fig:mainetti}
\end{figure}

Following the partial or complete destruction of the star is the formation of two tails of tidally stripped debris, a.k.a.~the ``debris stream.'' Figure \ref{fig:mainetti} -- adapted from \citet{mainetti17} -- gives an illustration. 
If the only force acting on the gas is the gravitational field of the SMBH, then we can determine the fallback rate -- the rate at which bound tidally stripped debris returns to the SMBH -- by following the procedure outlined by \citet{rees88}: the $i$$^{\rm th}$ fluid element in the debris stream has specific energy $\epsilon_{\rm i}$ and orbital period $T_{\rm i}$
\begin{equation}
\epsilon_{\rm i} = \frac{1}{2}v_{\rm i}^2-\frac{GM}{r_{\rm i}}, \quad T_{\rm i} = 2\pi \frac{a^{3/2}}{\sqrt{GM}} = 
\frac{\pi GM_{\bullet}}{\sqrt{2}}\left(-\epsilon_{\rm i}\right)^{-3/2}. \label{epscons}
\end{equation}
The total mass $M$ that has returned to the SMBH up to some orbital period $T$ is the sum over all the fluid elements that have energies $\epsilon_{\rm i} < \epsilon$, where $\epsilon$ is the energy associated with the orbital period $T$. $M$ is only a function of $\epsilon$, and hence the mass to have returned in time $T$ is $M(\epsilon[T])$ with $\epsilon[T]$ given by the inverse of Equation \eqref{epscons}, i.e.,
\begin{equation}
\epsilon[T] = -\left(\frac{\pi GM_{\bullet}}{\sqrt{2}}\right)^{2/3}T^{-2/3}. \label{epsofT}
\end{equation}
The rate at which material returns to pericenter can be determined by taking the difference in $M$ at times $T$ and $T+\Delta T$, which correspond to energies $\epsilon[T]$ and $\epsilon[T+\Delta T] \equiv \epsilon[T]+\Delta\epsilon$:
\begin{equation}
\Delta M = M(T+\Delta T)-M(T) = M(\epsilon+\Delta \epsilon)-M(\epsilon) \simeq \frac{dM}{d\epsilon}\Delta\epsilon+\mathcal{O}\left[\Delta\epsilon^2\right] = \frac{dM}{d\epsilon}\frac{d\epsilon}{dT}\Delta T
\end{equation}
Here we used a Taylor series to expand $M(\epsilon+\Delta\epsilon)$ in terms of $dM/d\epsilon$ and $\Delta \epsilon = (d\epsilon/dT)\Delta T$. The time rate of change of the supply of mass to the SMBH is therefore

\begin{equation}
\dot{M} = \frac{dM}{dT} = \frac{d\epsilon}{dT}\frac{dM}{d\epsilon} = \frac{2}{3}\left(\frac{\pi GM_{\bullet}}{\sqrt{2}}\right)^{2/3}T^{-5/3}\frac{dM}{d\epsilon}. \label{dMdT}
\end{equation}
In the last equality we used Equation \eqref{epsofT} to calculate the derivative $dM/d\epsilon$ as a function of time. Equation \eqref{dMdT} is our expression for the fallback rate from a TDE: the rate at which tidally destroyed material returns to pericenter. 

The function $M(\epsilon)$ is the cumulative distribution of mass as a function of energy: we sum the particles in successive energy bins from the most-bound to most-unbound fluid element. Depending on the stellar properties, this function (and $dM/d\epsilon$) will have a nontrivial energy dependence, and the early-time evolution of the rate of return of tidally disrupted debris to the SMBH will be variable. Indeed, if there is vanishingly small mass at the edge of our star, then $dM/d\epsilon$ for the most-bound orbit must be zero (and hence so is $dM/dT$), and the fallback rate must rise with time as the most-bound debris starts to return. However, as $T\rightarrow \infty$, material is returning from $\epsilon \simeq 0$, as shown explicitly in Equation \eqref{epsofT}, and hence $dM/d\epsilon(T\rightarrow\infty)\simeq dM/d\epsilon(\epsilon \simeq 0)$. Therefore, independent of the functional form of $dM/d\epsilon$, the asymptotic fallback rate is\footnote{If there is zero mass at the marginally bound radius, i.e., $dM/d\epsilon\bigg{|}_{\epsilon = 0}\equiv 0$, then $dM/d\epsilon \simeq d^2M/d\epsilon^2\bigg{|}_{\epsilon = 0}\epsilon$, and the fallback rate is $\dot{M} \propto T^{-7/3}$; see \citet{cufari22} for additional discussion.}
\begin{equation}
\textrm{[Complete disruptions:]          }\lim_{T\rightarrow \infty} \dot{M} = \frac{2}{3}\left(\frac{\pi GM_{\bullet}}{\sqrt{2}}\right)^{2/3}T^{-5/3}\frac{dM}{d\epsilon}\bigg{|}_{\epsilon = 0} \propto T^{-5/3}. \label{Mdotasy}
\end{equation}
At late times the fallback rate therefore scales as $\dot{M} \propto T^{-5/3}$. Although the dissipation that follows the fallback and the production of luminous emission is a far more intricate and complicated subject (see Section \ref{sec:luminosity_production} below), this simple power-law scaling remains one of the hallmark signatures of a TDE\footnote{It is unfortunate that the original derivation of this by \citet{rees88} contained an erroneous exponent of $-5/2$ instead of $-5/3$; this was corrected by \citet{phinney89}}.

Equations \eqref{dMdT} and \eqref{Mdotasy} are valid if the specific Keplerian energy is conserved, which requires that the only force acting on the debris be due to the gravitational field of the SMBH. Nonetheless, there are at least two other forces that the frozen-in approximation ignores, being pressure and self-gravity, and the latter is certainly not ignorable in the case of a partial disruption where a fraction of the star survives the tidal encounter (hence the ``complete disruptions'' prepended to Equation \ref{Mdotasy}). Specifically, in a partial TDE the surviving core exerts a gravitational influence, produces a time-dependent gravitational potential, and invalidates the conservation of specific Keplerian energy. 

Partial TDEs and their fallback rates can be understood with hydrodynamical simulations, among the first of which were \citet{khokhlov93a, guillochon13} and those that we discuss in the next section. However, the asymptotic fallback rate for partial TDEs, i.e., the analog of Equation \eqref{Mdotasy}, can be derived by noting that \footnote{We note that the late-time fallback rate of $\dot{M}\propto T^{-2.2}$ was inferred numerically for some stars by \citet{guillochon13} and prior to the analytical derivation presented here, despite their usage of the Keplerian energy formalism; see specifically their Figure 7.}: the debris in a partial TDE (or a complete disruption, for that matter) moves effectively radially from the SMBH, and we can analyze the one-dimensional motion of the debris under the combined gravitational influence of the SMBH and the surviving core and use the fact that the integrated mass between any fluid element and the SMBH is conserved to derive the fallback rate. The result is that $\dot{M}\propto T^{n_{\infty}}$, where\footnote{The factor of $17/(2\sqrt{73})\simeq 0.690$ in Equation 15 of \citet{coughlin19} is in error, and stems from the fact that the coefficient of $3/8$ in their Equation 14 should be $1/3$; the correct result is given by Equation \eqref{ninfty} here.} \citep{coughlin19}
\begin{equation}
n_{\infty} = -\frac{1}{6}\left(5+\sqrt{73}\right)-\frac{8}{\sqrt{73}}\left(\frac{1}{3}\frac{M_{\rm c}}{M_{\bullet}}\right)^{1/3}+\mathcal{O}\left[\left(\frac{M_{\rm c}}{M_{\bullet}}\right)^{2/3}\right] \simeq -2.257-0.649\left(\frac{M_{\rm c}}{M_{\bullet}}\right)^{1/3}. \label{ninfty} 
\end{equation}
Here $M_{\rm c} \simeq M_{\star}$ is the mass of the surviving stellar core. 

There are two surprising features of Equation \eqref{ninfty}, the first of which is that the dependence on the core mass only enters as its ratio to the mass of the SMBH, which in general is $\lesssim 10^{-5}$ (though this becomes more important when the mass of the disrupting object becomes comparable to the mass of the star; see \citealt{wang21, kremer23}). The second term in Equation \eqref{ninfty} therefore introduces corrections that are at the $\sim 0.1\%$ level, and if we recognize that $-2.257 \simeq -9/4$, we find
\begin{equation}
\textrm{[Partial disruptions:]         }\lim_{T\rightarrow \infty}\dot{M} \propto T^{-9/4}. \label{Mdotasypartial}
\end{equation}
This decline with time is significantly steeper than the $\propto T^{-5/3}$ falloff for complete disruptions given in Equation \eqref{Mdotasy} and is (effectively) independent of the mass of the surviving core, and hence the original star. The second surprise is that setting $M_{\rm c} = 0$ in Equation \eqref{ninfty}, which should correspond to complete disruption, does not yield the complete disruption fallback rate of $\dot{M} \propto T^{-5/3}$. The meaning of this discrepancy is subtle and also reveals the physical origin of the steeper falloff in partial TDEs: at asymptotically late times, the material returning to the SMBH in a partial TDE originates from the Hill sphere of the surviving core, which delineates the boundary between material that accretes onto the SMBH vs.~the surviving core. The mass near the Hill sphere is drained much more rapidly than in the core-less case of a complete disruption, as the core ``steals'' mass that would otherwise accrete onto the SMBH. The Hill sphere exists in the limit that $M_{\rm c} \rightarrow 0$, and hence even a vanishingly small core mass will eventually cause the fallback rate to steepen to $\propto T^{-9/4}$ (see the discussion in \citealt{miles20}). This is not the same as setting the core mass to identically zero; redoing the analysis that leads to Equation \eqref{Mdotasypartial} with $\mu \equiv 0$ yields $\dot{M} \propto T^{-5/3}$ \citep{coughlin19}.

The models discussed here are approximate in that they neglect complicating geometrical and physical effects, and a number of efforts have elucidated the accuracy of the preceding predictions with the aid of hydrodynamical simulations. We discuss these efforts below, but we first briefly address the perturbative limit for completeness.

\subsection{The perturbative limit: $\beta \ll 1$}
\label{sec:perturbative}
When $\beta \ll 1$, the tidal field perturbs the star slightly, and one linearizes the fluid equations to yield a set of coupled, first-order, partial differential equations. The solutions to these equations are in the form of eigenmodes, which can be thought of as standing waves excited within the star, that describe the oscillatory response of the star to the tidal field. The energy imparted into these oscillations comes at the expense of the orbital energy of the star, which leads to the ``capture'' of the star by the perturbing object; we refer the reader to, e.g., \citet{fabian75, press77, lee86, ogilvie14} for derivations, astrophysical applications, and reviews of this limit. Here we will focus on the case more relevant to TDEs, where $\beta \sim 1$ (i.e., where the star can actually be destroyed by tides).

\subsection{The mildly disruptive limit: $\beta \sim 1$}
\label{sec:mildly_disruptive}
As $\beta$ nears one, the forces due to self-gravity, pressure, and tides are all of comparable magnitude, and for this reason the mildly destructive limit has been modeled largely numerically -- but supplemented by the frozen-in approximation described in Section \ref{sec:frozen} above -- with the result being that the star can either be completely or partially destroyed by tides. \citet{nolthenius82} and \citet{evans89} performed some of the first smoothed-particle hydrodynamics (SPH) simulations of the complete disruption of a solar-like, 5/3-polytrope with $\beta = 1$. A 5/3-polytrope refers to the type of star -- a polytropic relationship equates the pressure $p$ of a fluid element to its density $\rho$ via $p = K\rho^{\gamma}$, with $K$ and $\gamma$ constants; a $5/3$-polytrope has $\gamma = 5/3$ and is ``solar-like'' if its mass and radius are those of our sun. SPH \citep{lucy77, gingold77} is a finite-mass numerical algorithm for solving the fluid equations; the interested reader can consult, e.g., \citet{price12} for a review. \citet{evans89} also performed finite-volume simulations, finding that the two methods agreed very well with one another. 

\begin{figure}[htbp] 
   \centering
   \includegraphics[width=0.55\textwidth]{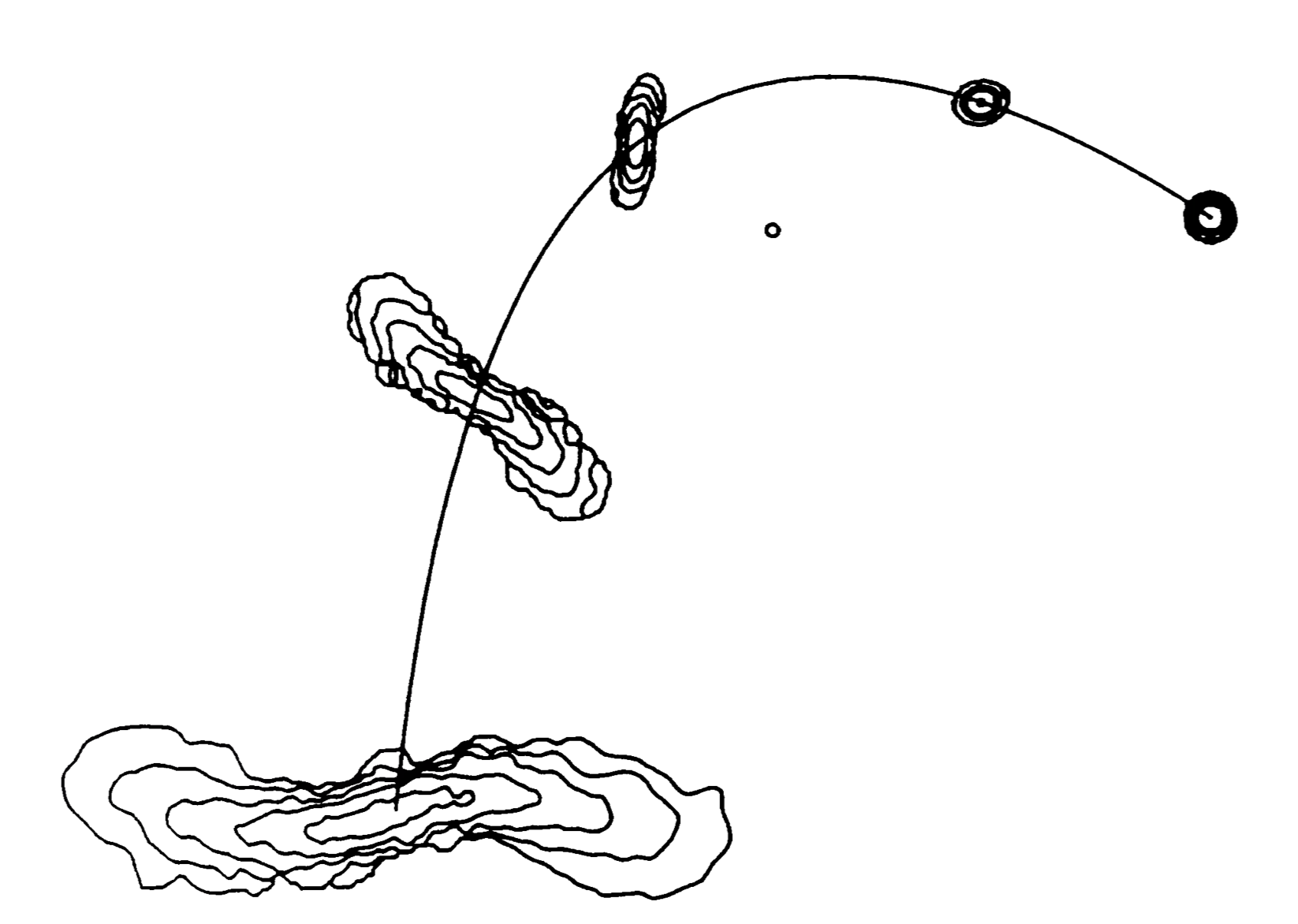} 
   \includegraphics[width=0.445\textwidth]{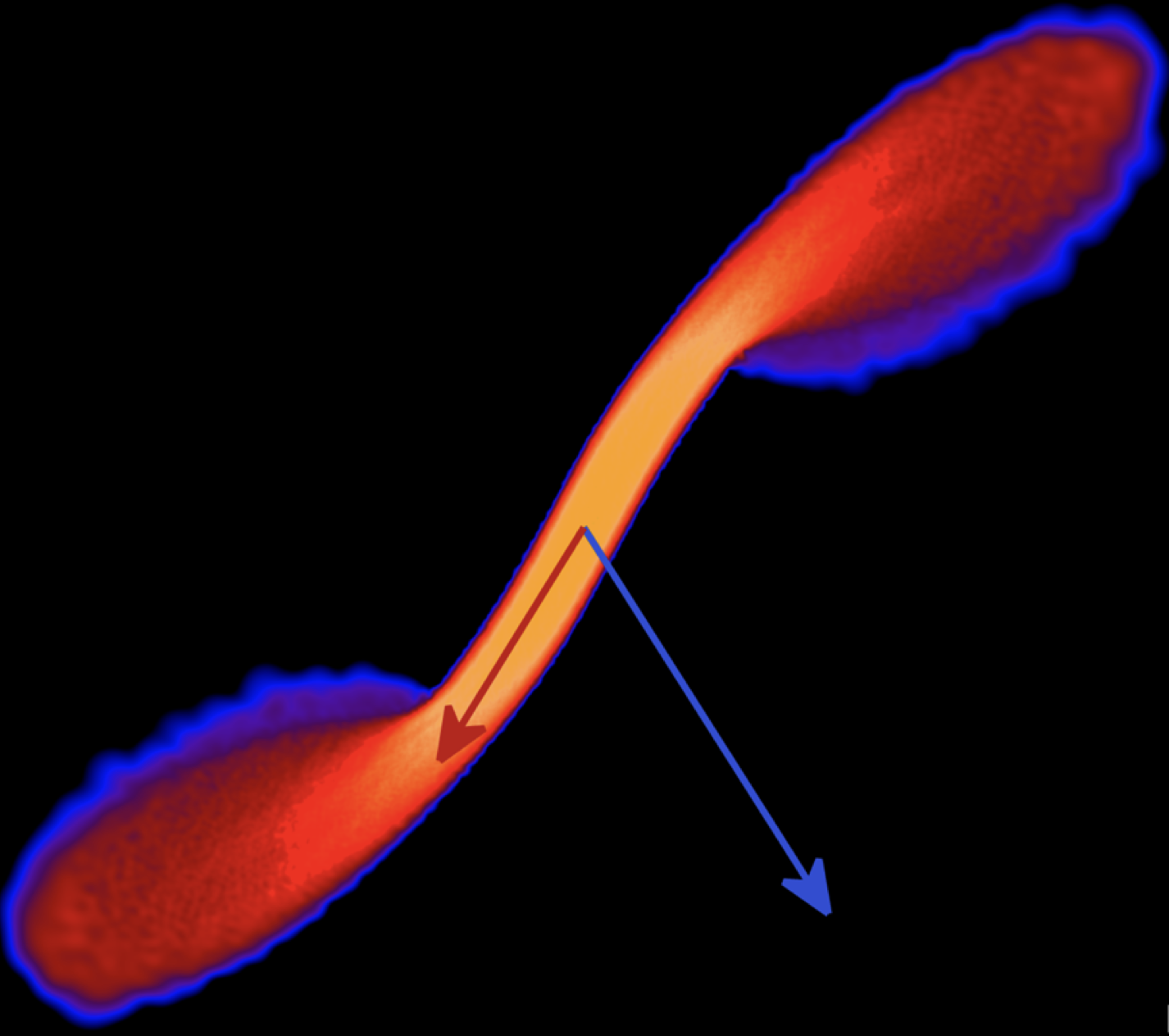}
   \caption{{\bf Left:}  one of the first SPH simulations of a TDE, performed by    \citet{evans89}, who used $4\times 10^{4}$ SPH particles and who also performed grid-based simulations. The star is modeled as a 5/3-polytrope and enters from the top-right, the point of closest approach has $r_{\rm p} = r_{\rm t}$, and the circle indicates the Schwarzschild radius of the SMBH that has a mass of $10^6M_{\odot}$ (note that \citealt{evans89} increased the spatial scale by a factor of 15 to aid in the presentability of the results). {\bf Right:}  the same simulation as in the left in terms of physical setup (i.e., $5/3$-polytrope, $10^6 M_{\odot}$ SMBH, $\beta = 1$) but with $10^6$ particles, showing the same tidally disrupted debris stream has formed as a byproduct of the encounter; adapted from \citet{bonnerot17}.}
   \label{fig:first_disruptions}
\end{figure}

The left panel of Figure \ref{fig:first_disruptions} shows isodensity contours from a $4\times 10^{4}$-particle simulation described in \citet{evans89} with $\beta = 1$: the star is distorted from its spherical shape as it nears pericenter, after which the debris spreads differentially with time and the star is destroyed. Since these seminal works, the destruction of a 5/3-polytropic star by a $10^6M_{\odot}$ SMBH has effectively become the prototypical TDE, and is the encounter for which the community has reached a consensus as concerns the initial evolution of the tidal debris. The right panel of Figure \ref{fig:first_disruptions} gives a more recent simulation of the same disruption from \citet{bonnerot17}. Works that considered $5/3$-polytropes in the $\beta \sim 1$ regime, listed chronologically by year and within that year alphabetically, include \citet{khokhlov93a, laguna93, fulbright95, ayal00, ivanov01, bogdanovic04, gomboc05, lodato09, guillochon13, hayasaki13, coughlin15, gafton15, hayasaki16, bonnerot17, mainetti17, wu18, golightly19b, steinberg19, miles20, park20, norman21, cufari22}. 

\begin{figure}
\centering
    \includegraphics[width=0.5\textwidth]{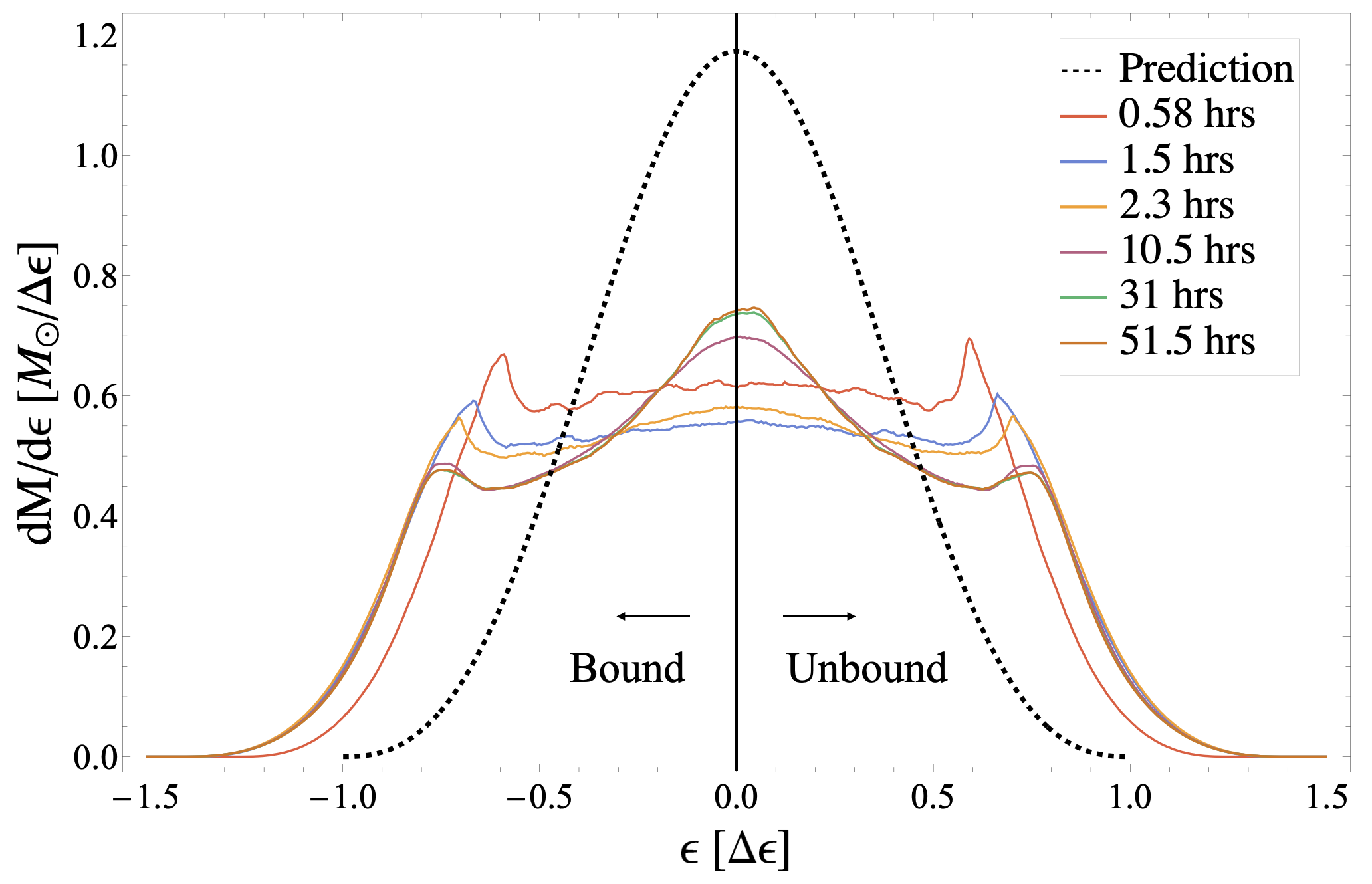}
    \caption{The $dM/d\epsilon$ curve from the disruption of a $5/3$-polytropic and solar-like star by a $10^6M_{\odot}$ SMBH, numerically simulated with the SPH method with $\sim 10^8$ particles. The black, dashed curve is the frozen-in prediction for this specific type of star, as described in \citet{lodato09}. The overall spread in the energy ($\pm \Delta \epsilon$) is in good agreement with the frozen-in prediction, but the temporal evolution of the curve and the large discrepancies with the analytic model suggest that additional effects are important. Figure adapted from \citet{fancher23}, with time post-pericenter indicated in the legend. }
    \label{fig:fancher_dmde}
\end{figure}

As noted in Section \ref{sec:frozen}, one of the predictions of the frozen-in model is that there is a spread in the Keplerian energies imparted to the tidal debris, $\Delta\epsilon = \epsilon_{\star}\left(M_{\bullet}/M_{\star}\right)^{1/3}$, and this can be measured from simulations. Figure \ref{fig:fancher_dmde} gives one such example (using $\sim 10^8$ SPH particles) that adopted the same physical parameters as in \citet{evans89} and is adapted from \citet{fancher23} (the original simulations were presented in \citealt{norman21}), where the colored curves represent time post-pericenter, as shown in the legend. We see that the overall energy spread is in broad agreement with the frozen-in prediction, but there are some notable discrepancies, especially with the prediction that follows from the frozen-in model applied to this specific star (dashed curve; see \citealt{lodato09}; note also that their analytic curve in Figure 6 was arbitrarily renormalized to match their numerically obtained peak). It is also apparent that the curve evolves with time, i.e., the energy spread is not frozen in, and this is due to the self-gravitating nature of the debris  (\citealt{steinberg19}; see also \citealt{fancher23}, specifically their Figure 2). 

\begin{figure}
    \includegraphics[width=0.425\textwidth]{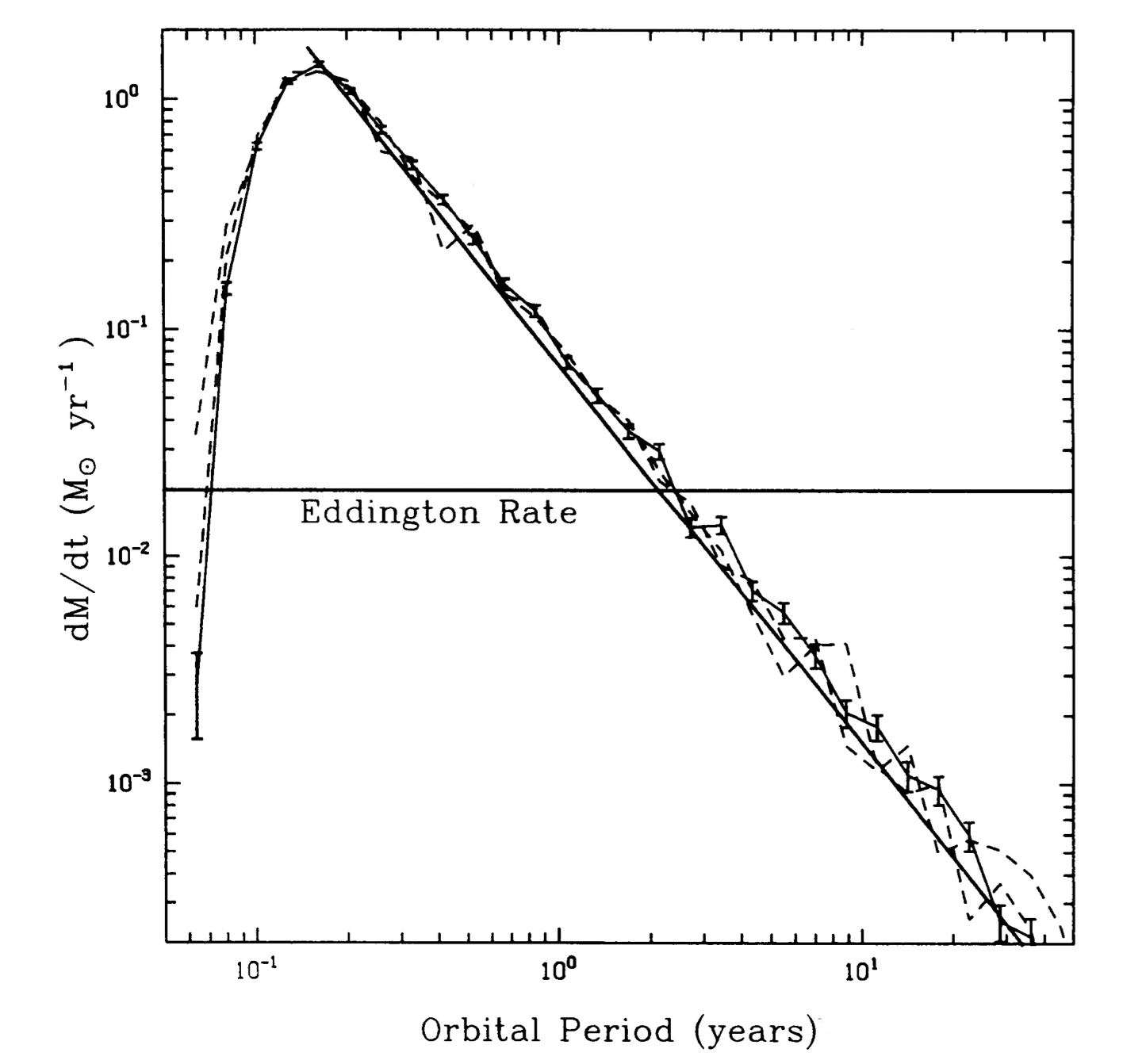}
     \includegraphics[width=0.575\textwidth]{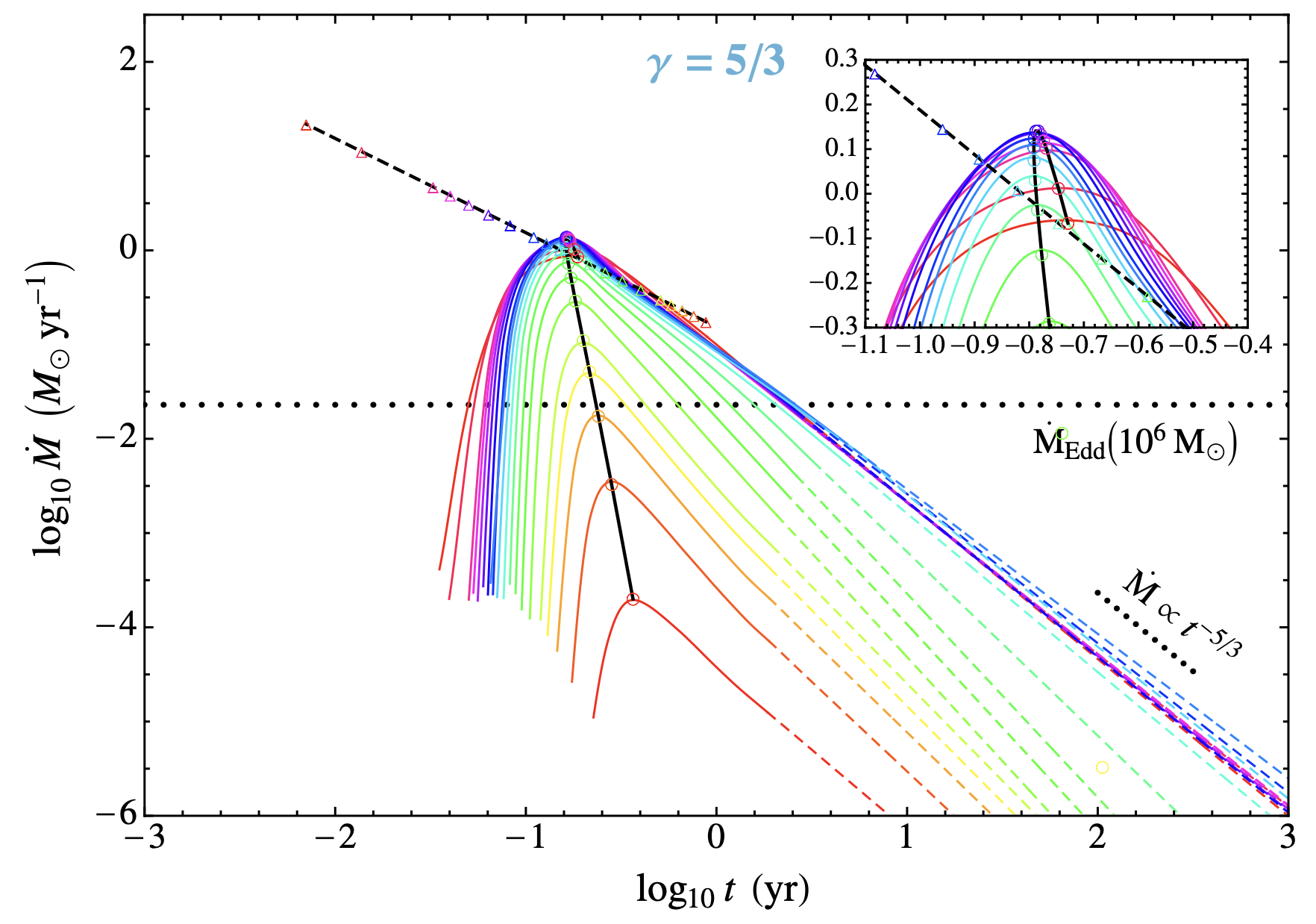}
     \caption{{\bf Left:}  the fallback rate of material measured from the simulations performed in \citet{evans89}, with the solid and straight line indicating the $\propto t^{-5/3}$ fallback rate (different curves show different resolutions). {\bf Right:}  fallback rates from the simulations performed in \citet{guillochon13} for the same type of star and same SMBH mass as the left panel. Different curves correspond to different values of $\beta$ (the one corresponding to the left is the dark-blue curve, shown most clearly in the inset).}
     \label{fig:dmdt_evans_guillochon}
\end{figure}

The fallback rate $\dot{M}$ -- the rate at which tidally destroyed material returns to the SMBH -- can also be deduced from simulations. Figure \ref{fig:dmdt_evans_guillochon} shows the fallback rate measured from the simulations performed by \citet{evans89}. The solid line in this figure is the $\propto T^{-5/3}$ prediction. The right panel of this figure shows the fallback rate measured from the simulations in \citet{guillochon13}, where the blue and dark-red curves correspond to complete disruptions (the dark blue curves in the inset closely mimic the simulation on the left and have $\beta \sim 1$). The results agree well with one another, and fairly well with the frozen-in approximation -- the latter predicts that the most-bound debris returns $\sim 40$ days after disruption \citep{lodato09}, whereas the numerical results shown in Figure \ref{fig:dmdt_evans_guillochon} find a return time of $\sim 30$ days.

In addition to complete disruptions, $\beta \sim 1$ encounters can -- as mentioned in Section \ref{sec:frozen} -- result in partial disruptions, in which a fraction of the star survives the tidal encounter. The curves with smaller peak magnitudes in the right panel of Figure \ref{fig:dmdt_evans_guillochon} (those colored dark red, orange, and green) represent the fallback rates from partial disruptions of a $5/3$ polytrope, where the curve with the lowest peak (at which almost no mass loss occurs) has the smallest value of $\beta = 0.5$. The peak value of the fallback rate and the time at which the peak occurs depend on $\beta$ when the disruption is only partial, with the time to reach the peak generally occurring at later times as $\beta$ decreases; \citet{guillochon13} find that the time to peak (peak value) is $\sim 50$ days ($\sim 1 M_{\odot}$ yr$^{-1}$ or $\sim 6.3$ kg s$^{-1}$) for the complete disruption of a $5/3$ polytrope, and $\sim 100$ days ($\sim 10^{-2} M_{\odot}$ yr$^{-1}$ or $\sim 0.0063$ kg s$^{-1}$) for $\beta = 0.55$ (see their Figure 12 and their Appendix). Many of the papers mentioned in the previous paragraph studied the transition from partial to complete disruption of the polytropic star, with the result being that a $5/3$-polytrope is partially destroyed for $\beta \simeq 0.55$ and completely destroyed for $\beta \simeq 0.9$; see, e.g., Figure \ref{fig:mainetti} above. 

The Universe creates many more types of SMBH than those with $10^6 M_{\odot}$ and many more types of star than $5/3$-polytropes, and to more broadly assess our understanding of TDEs we need to analyze their parameter space: for a given SMBH mass and type of star, what is the outcome of the tidal interaction as a function of the point of closest approach to the SMBH? What are the effects of the equation of state of the gas? How do general relativistic effects modify the outcome relative to a Newtonian description?

We have started to explore this parameter space. \citet{khokhlov93a} considered the disruption of three different polytropes, finding that the point of closest approach of a $\gamma = 4/3$ polytrope ($p \propto \rho^{4/3}$) needed to be $\sim 0.6\times r_{\rm t}$ ($\beta \simeq 1.6$) for the star to be completely destroyed, and that stars with smaller polytropic indices -- which have higher central densities compared to their average densities -- required closer pericenter distances to be completely destroyed. \citet{guillochon13} performed a dense grid of simulations of $\gamma = 5/3$ and $\gamma = 4/3$, finding that there is a strong (weak) dependence on the peak fallback rate properties with $\beta$ for partial (complete) disruptions, and also that the fallback rate declines more rapidly with time for partials. There are many more works that focused on the impact of stellar type (e.g., \citealt{lodato09, mainetti17, golightly19b, lawsmith19, lawsmith20, ryu20a, jankovic23b, sharma24}) and stellar spin (e.g., \citealt{golightly19a, sacchi19}), relativistic effects (e.g., \citealt{beloborodov92, kesden12, sadowski16, gafton19, ryu20d, andalman22}), the stellar orbit (e.g., \citealt{hayasaki13, clerici20, park20, cufari22}), and black hole binarity (e.g., \citealt{chen08, ricarte16, coughlin17, vigneron18, mockler23, melchor24}).

\subsection{The highly destructive limit: $\beta \gg 1$}
\label{sec:highly_destructive}
\begin{figure}[htbp] 
   \centering
   \includegraphics[width=0.495\textwidth]{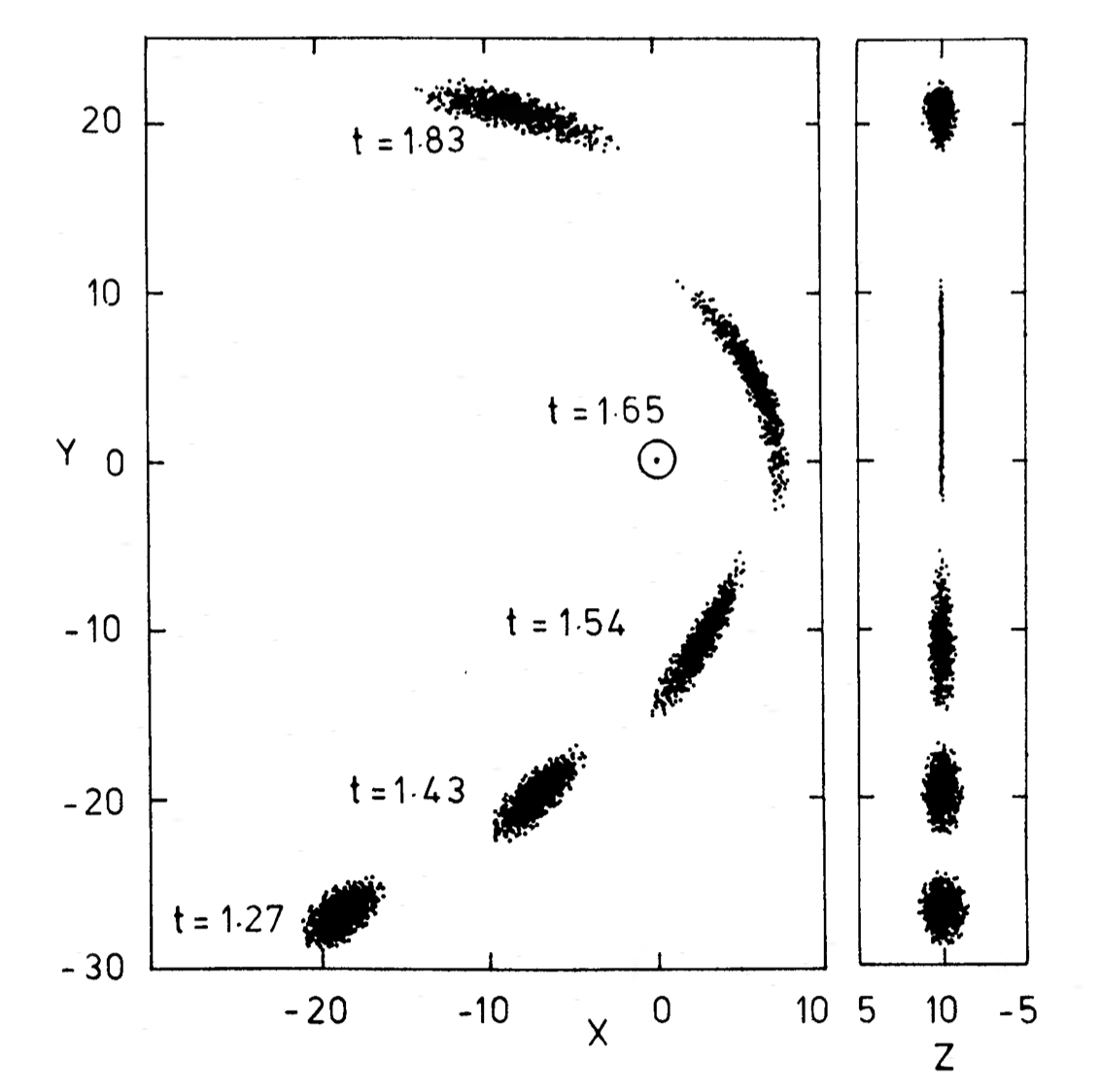} 
 \includegraphics[width=0.485\textwidth]{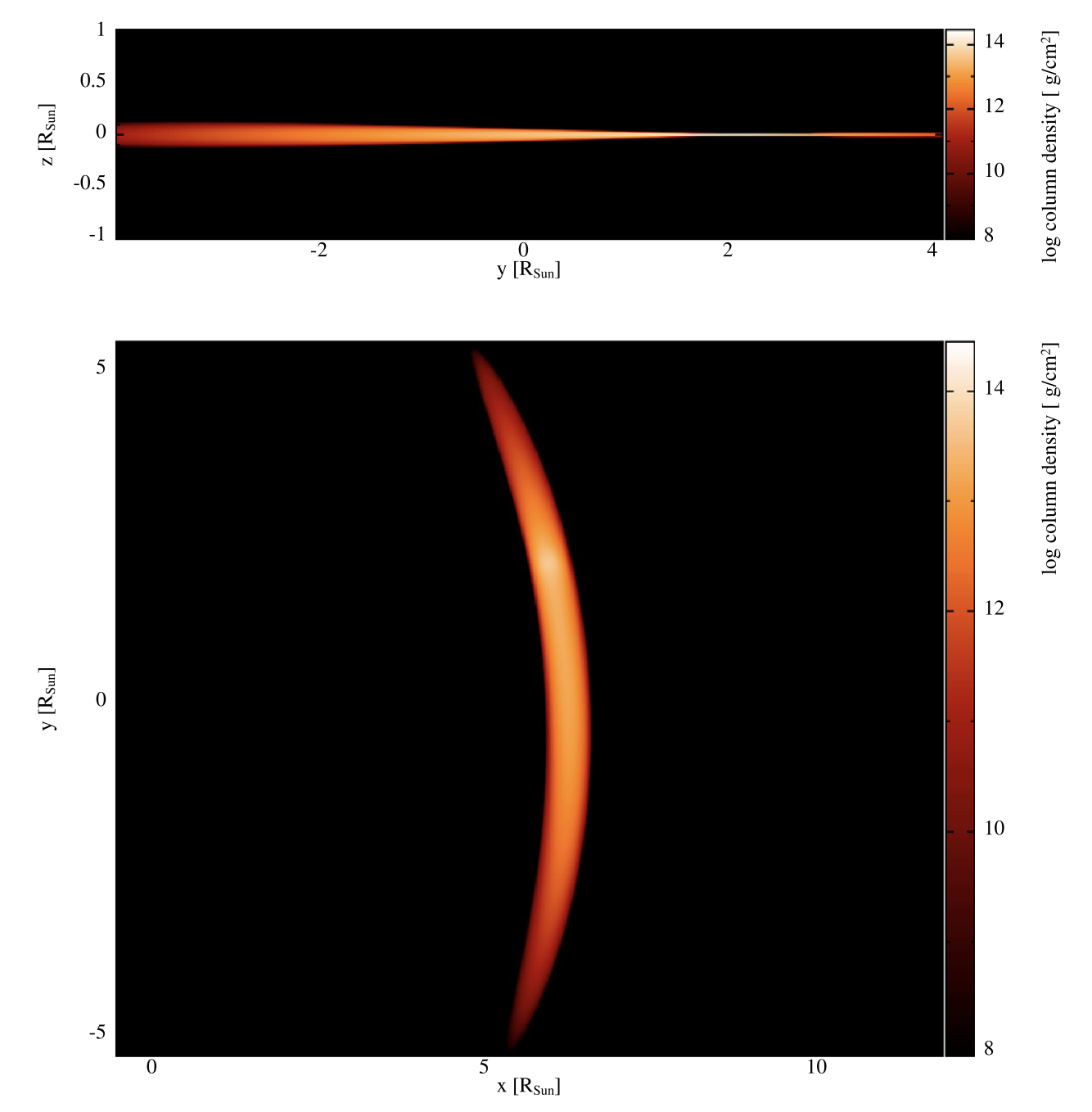} 
   \caption{{\bf Left:}  one of the first SPH simulations of a TDE by \citet{bicknell83}, for which $\beta = 15.6$ (i.e., pericenter distance $r_{\rm p} = r_{\rm t}/15.6$), the star is modeled as a $\gamma = 5/3$ polytrope, and the SMBH mass is $M_{\bullet} = 10^{5}M_{\odot}$. The distributions of particles onto the orbital plane of the star ($Y-X$) and out of the plane ($Y-Z$) are shown, and for this simulation 500 or 2000 particles were used. The circle shows the Schwarzschild radius of the SMBH. {\bf Right:}  the in-plane and out-of-plane projections of the density from an SPH simulation of a TDE by \citet{norman21} with $\beta = 16$ (i.e., comparable to the left panel) for a 5/3-polytrope and a $10^6M_{\odot}$ SMBH. Here $\sim 10^8$ particles were used. In both cases the star is stretched into a crescent shape within the plane and compressed into a small fraction of its original radius in the vertical direction. Figure adapted from \citealt{coughlin22a}.}
   \label{fig:crushed}
\end{figure}

When $\beta \gg 1$, tides are more than capable of destroying the star, but this regime is qualitatively distinct from the $\beta \sim 1$ (and full-disruption) regime for two reasons. First, and as pointed out by \citet{carter82}, the out-of-plane compression that the star experiences when $\beta \gg 1$ becomes extreme. If one assumes that the compression is adiabatic and that pressure can be ignored until the star is maximally compressed, then the density of the star increases by a factor of $\sim \beta^3$ (assuming a $\gamma = 5/3$ equation of state) and the temperature by $\propto \beta^2$, potentially leading to a thermonuclear ``pancake detonation'' of the star \citep{carter83}. \citet{bicknell83}, however, claimed that these scalings are not upheld and that shocks halt the compression, and performed some of the first SPH simulations (with $\lesssim 2000$ particles) of TDEs to substantiate their claim. They found that the star was highly compressed with increasing $\beta$ (see the left panel of Figure \ref{fig:crushed}), but at the expense of generating large amounts of entropy (see their Figure 7) that they attributed to shocks.
The ``deep TDE'' problem has since been revisited by many (e.g., \citealt{carter83, luminet85, luminet86, luminet89, laguna93, gomboc05, brassart08, brassart10, stone13, sadowski16, tejeda17, darbha19, gafton19, norman21, coughlin22a, kundu22, ryu23}), and it was determined by \citet{coughlin22a} that neither \citet{carter82} nor \citet{bicknell83} were completely correct: the density increases more weakly than $\propto \beta^3$, but simply because the pressure rises with the density and resists the tidal compression throughout the encounter. They also showed that shocks formed during the compression of the star are weak, reaching a Mach number of at most $\sim 1.5$ for $\beta = 16$, and concluded that much of the heating observed by \citet{bicknell83} was spurious and numerical (see also Figure 5 of \citealt{norman21}). 

\begin{figure}[htbp] 
   \centering
   \includegraphics[width=0.475\textwidth]{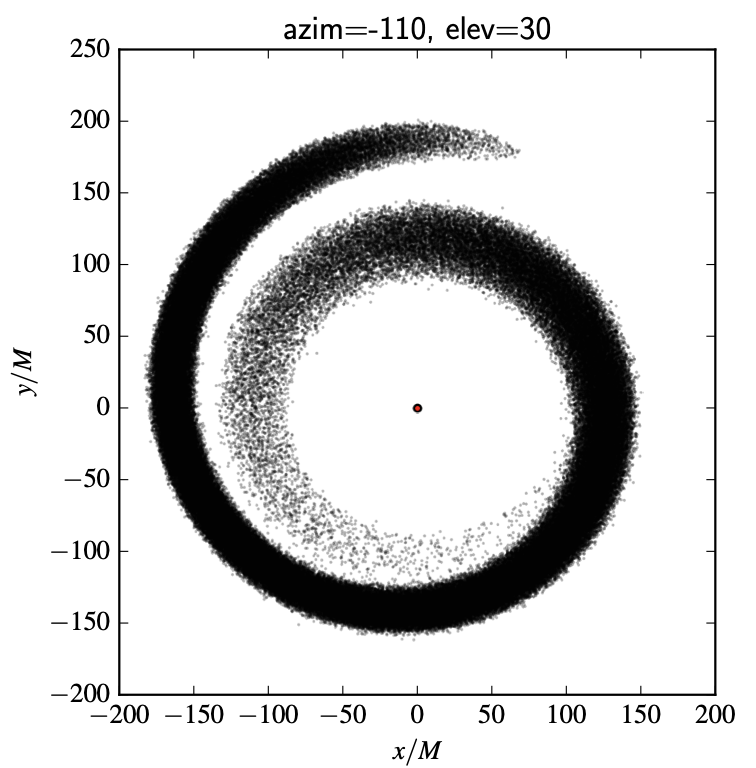} 
   \includegraphics[width=0.455\textwidth]{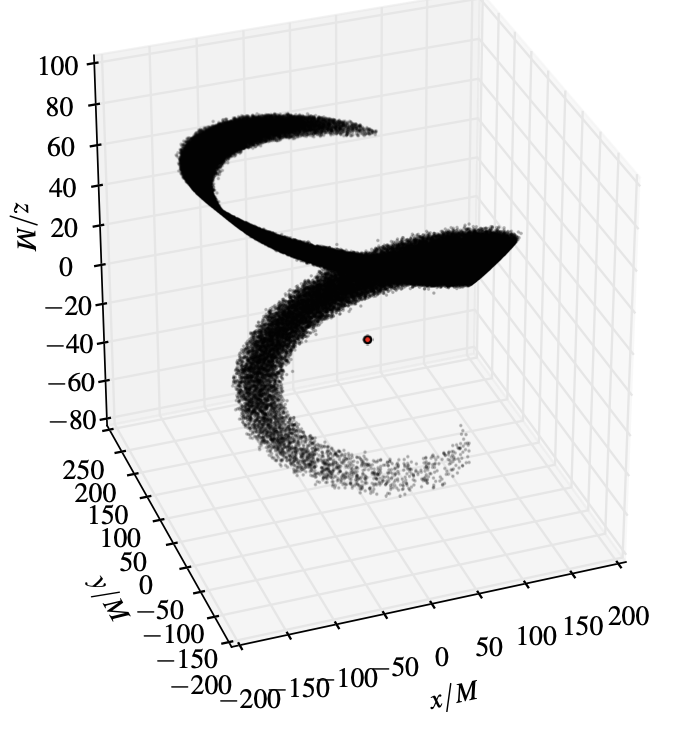} 
   \caption{The morphology of the tidally disrupted debris from the destruction of a star by a rotating black hole, where the left is the in-plane projection and the right is a three-dimensional representation. The pericenter distance in this case was $\sim 1.5 \, GM_{\bullet}/c^2$, which results in substantial in-plane and out-of-plane precession and the dramatic stretching of the star. Figure adapted from \citet{tejeda17}}
   \label{fig:tejeda}
\end{figure}

The second reason deep TDEs are phenomenologically distinct is because of strong-gravity effects: the ``point of no return'' for an object orbiting a SMBH coincides with the direct capture radius, $r_{\rm dc}$, which is $r_{\rm dc} = 4GM_{\bullet}/c^2$ -- twice the Schwarzschild radius -- for a spin-zero SMBH and a parabolic orbit. The ratio of $r_{\rm p}$ (the pericenter distance) to $r_{\rm dc}$ in this case is then 
\begin{equation}
\frac{r_{\rm p}}{4GM_{\bullet}/c^2} = \frac{1}{\beta}\frac{R_{\star}}{4GM_{\star}/c^2}\left(\frac{M_{\bullet}}{M_{\star}}\right)^{-2/3} \simeq \frac{11.8}{\beta}\left(\frac{R_{\star}}{R_{\odot}}\right)\left(\frac{M_{\star}}{M_{\odot}}\right)^{-1}\left(\frac{M_{\bullet}/M_{\star}}{10^{6}}\right)^{-2/3}. \label{rprdcS}
\end{equation}
For typical numbers ($R_{\star} = R_{\odot}$, $M_{\star} = M_{\odot}$, $M_{\bullet} = 10^6 M_{\odot}$), direct capture occurs for $\beta \gtrsim 11.8$. Any TDE that produces observable emission and has $\beta > 11.8$ therefore must have been disrupted by a spinning SMBH; for maximal SMBH spin and a prograde stellar orbit, $r_{\rm dc} = GM_{\bullet}/c^2$, and in this case the maximum $\beta$ is
\begin{equation}
\beta_{\rm max} \simeq 47.1\left(\frac{R_{\star}}{R_{\odot}}\right)\left(\frac{M_{\star}}{M_{\odot}}\right)^{-1}\left(\frac{M_{\bullet}/M_{\star}}{10^6}\right)^{-2/3}. \label{betamax}
\end{equation}
A related quantity is the ``Hills mass'': setting $\beta_{\rm max} = 1$ in the previous expression implies that the direct capture radius coincides with the tidal radius, and rearranging the resulting equation for the black hole mass and adopting solar values for the star gives
\begin{equation}
    M_{\rm Hills} = 3\times 10^{8}M_{\odot}. \label{hillsmass}
\end{equation}
Since a star is $\sim$ destroyed at $r_{\rm t}$, Equation \eqref{hillsmass} is the maximum-mass SMBH that is capable of destroying a (solar-like) star. The direct capture radius is characterized by a diverging relativistic advance of periapsis angle -- in general relativity orbits do not close, and this can be characterized by the additional angle through which the pericenter advances on a per-orbit basis, a.k.a.~the general relativistic advance of periapsis. At $r_{\rm dc}$ this angle diverges, meaning that an object that reaches $r_{\rm dc}$ would orbit the black hole (at $r_{\rm dc}$) indefinitely. This implies that as $\beta$ approaches the direct capture value\footnote{It should be noted that the likelihood of attaining small pericenter distances without being directly captured is small, not only because of the smaller geometrical region of parameter space in which such high-$\beta$ encounters reside, but also because of general relativistic effects. For SMBHs with near-maximal spin, the cumulative distribution function of $\beta$ declines as $\propto \beta^{-7/3}$ \citep{coughlin22b} for $\beta \gtrsim 10$ (see Figure 3 of \citealt{coughlin22b}), e.g., the total number of events with $\beta > 40$ is $\sim 4\%$ the number with $\beta > 10$ (which itself is a small subset of all events with $\beta > 1$).}, the star is stretched by increasingly large factors before exiting the tidal sphere. Figure \ref{fig:tejeda} shows an example of such an encounter, where the stellar debris ``winds around'' the SMBH multiple times before receding to larger distances. 

Our understanding of the disruption process, the formation of a stream of debris, and the return of that debris to pericenter, is therefore on relatively firm theoretical ground. Unfortunately, none of these processes creates anything in the way of observable emission\footnote{\citet{kasen10} suggested that a ``recombination transient'' is produced as the hot plasma within the debris stream cools and recombines, but more recent estimates by \citet{coughlin23} indicate that such a transient is almost certainly undetectable (except, perhaps, for extremely nearby and highly improbable TDEs).}, and to create light from the kinetic energy of the debris we need dissipation -- this is the subject of the next section. 


\section{Luminosity production and light curves}\label{sec:luminosity_production}
\subsection{How is luminosity produced?}\label{sec:howislum}

The observed flares from tidal disruption events have estimated peak bolometric luminosities of $\sim 10^{43}- 10^{45}$ erg/s \citep{Hung2018, mockler_weighing_2019, Hammerstein2023, Yao2023}\footnote{Largely calculated from optical and UV emission, and therefore possibly missing extreme UV or X-ray emission.}. While there are many uncertainties as to the exact emission mechanisms that produce the observed luminosity, we do know the luminosity ultimately comes from the dissipation of the kinetic energy (KE) of the gas orbiting the black hole. Very broadly, kinetic energy is converted into luminosity through gas hitting other gas – through shocks, stream collisions, and accretion.  Because of this, the efficiency of the conversion from mass to energy is dependent on the orbital velocity of the gas, and therefore on its distance from the black hole. 
Consider the limit that gas is moving at the speed of light, $c$, the orbital velocity that defines the gravitational radius, $R_g$. In this limit, all of the gas’s mass-energy budget is in kinetic energy, and you would theoretically approach a maximum efficiency of 1. 
Along these lines, we often write down the luminosity from accretion as an efficiency  ($\varepsilon$) multiplied by $\dot{M}$ multiplied by the speed of light squared:
\begin{equation}
    L = \varepsilon \dot{M} c^2
\end{equation}

The maximum possible efficiency at a given radius is then proportional to $v_{\rm orbit}^2/c^2 \propto R_g/r$. This makes black holes incredibly efficient engines, and is the reason why they (or more accurately the gas surrounding them) are some of the brightest objects in the universe. Simulations of accretion disks have found that the efficiency of conversion from mass to energy in an accretion disk is often close to $10\%$ ($\varepsilon = 0.1$), and observations have confirmed this \citep[e.g.][]{abramowicz_foundations_2013, davis_radiative_2011}. For comparison, nuclear burning of Hydrogen to Helium in the sun only converts $\sim 0.7\%$ ($\varepsilon = 0.007$) of the rest mass energy of Hydrogen into luminosity \citep{bethe_energy_1939}.  

After disruption, the gas from the star starts on very eccentric orbits with large semi-major axes. To produce an observable transient, the gas must interact and release energy (e.g., through interactions like those shown in Figure~\ref{fig:streamcollisions}). This same process will move gas from larger, higher energy orbits, onto smaller orbits and eventually allow it to accrete onto the black hole. If angular momentum is conserved while energy is lost, we can use the definition of the loss cone (see Equation~\ref{eq:losscone}) angular momentum to determine the radius of the circular orbit with the same angular momentum as the initial very eccentric orbit.

\begin{align}
     l_{\rm circ} &= \sqrt{G M_\bullet \; a} = \sqrt{G M_\bullet \; 2 r_p}  
\end{align}

Therefore, the circular orbit with approximately the same angular momentum as the initial nearly parabolic ($e\rightarrow 1$) orbit has a radius $= 2 R_p$ (or $2 R_t$ if the star is disrupted at the tidal radius). We define this radius as the ‘circularization radius’, and use it as an approximation for the radius of the accretion disk that will eventually form around the black hole. 

Ultimately, most of the energy will likely be released through accretion processes close to the black hole, because the efficiency is much higher than at larger radii, as explained above. However, it is not immediately obvious where most of the energy from the initial observed flare comes from (see Figure~\ref{fig:TDEefficiencies} for an analysis of observed efficiencies). In addition to requiring a lot of energy to be released, the initial flare also requires high luminosity (power), and so sufficient energy must be released over relatively short timescales (observed peak timescales are weeks - months). 

We want to constrain where the emission is coming from for multiple reasons, one of the most pressing being that we would like to use TDEs to help understand the efficiency of black hole growth. We discussed above how black holes are very efficient at converting mass to luminosity, but we also want to understand the closely connected question of how efficient they are at actually pulling mass in. If the mass accretion rate continues to grow, at some point the luminosity will become high enough that radiation pressure will become comparable to the gravitational pressure of accreting mass. At this point, the radiation will push back against accreting mass and reduce the efficiency of accretion. If we assume spherical symmetry (clearly not the best assumption if material is accreting in a disk, but illuminating nonetheless), we can calculate the point where the radiation force will perfectly balance the gravitational force on in-falling mass. 
This allows us to define a luminosity limit (the ``Eddington limit"), beyond which the efficiency of accretion will be significantly reduced. 

\begin{align}
    F_{\rm rad} & = F_{\rm grav} \\
    P_{\rm rad} \sigma & = F_{\rm grav} \\
    \frac{L_{\rm edd}}{4 \pi r^2 c} \sigma & = \frac{G M m}{r^2} \\
    L_{\rm edd} & = \frac{4 \pi c G M}{(\sigma / m)} \\
    L_{\rm edd} & = \frac{4 \pi c G M}{\kappa} 
\end{align}

Notably, the Eddington luminosity does not depend on the radius of the accreting object. Here the cross-section `$\sigma$' can be thought of as the fraction of radiation intercepted by the gas, which depends on the composition and ionization of the gas. For a given set of gas properties we can substitute `$\sigma /m$' (cross-section/particle mass) for the opacity `$\kappa$' of the gas. We often assume that the gas is mostly made up of ionized Hydrogen and use the Thomson electron scattering opacity ($ = \kappa_T$ ) to estimate the Eddington luminosity\footnote{Note that while the Thomson electron scattering opacity is probably a reasonable approximation for determining the order of magnitude radiation pressure, it does not include the very important contribution of absorption opacities that are key to determining the reprocessing of high-energy radiation and therefore the observed fraction of luminosity at different wavelengths (the spectral energy distribution).}. 

The mass fallback rates after stars are disrupted regularly reach super-Eddington values for canonical disk efficiencies of $\varepsilon = 0.1$, and so if material is able to form a disk during the initial flare, it will likely be fed near or above the Eddington limit for its black hole. Most AGN accrete well below the Eddington limit, and so observing and modeling TDEs is an important and promising way to study the limits of black hole accretion. However, to do this we need to know when the luminosity is being produced by a forming accretion disk, and when it is being produced by other processes such as stream collisions or ``nozzle shocks" \citep{evans89} that are not as directly connected to the growth rate of the black hole.

\subsection{Circularization shocks and disk formation}\label{sec:circularization}
There is considerable debate in the field as to what produces the optical and UV luminosity around the peak of the light curve. Where most of the initial flare's energy comes from depends on how quickly material can move from its very eccentric initial orbits to orbits of order the circularization radius. Shocks and accretion on the size scale of the circularization radius are significantly more efficient at converting mass to luminosity (as they have much larger KE reservoirs to draw from) than shocks at larger radii. Therefore, once material is orbiting the black hole at smaller radii, the majority of the emission should come from these radii (even if additional radiation is emitted at larger radii). If this process happens on a timescale shorter than, or of order the peak timescale, then most of the energy released in the observed flare will come from near the black hole from the process of forming an accretion disk. If this process takes longer than the observed peak timescale, the energy that goes into the peak of the flare must come from the gas interactions that circularize debris from orbits with initially large semi-major axes.  

\begin{figure}[htbp] 
   \centering
   \includegraphics[width=0.45\textwidth]{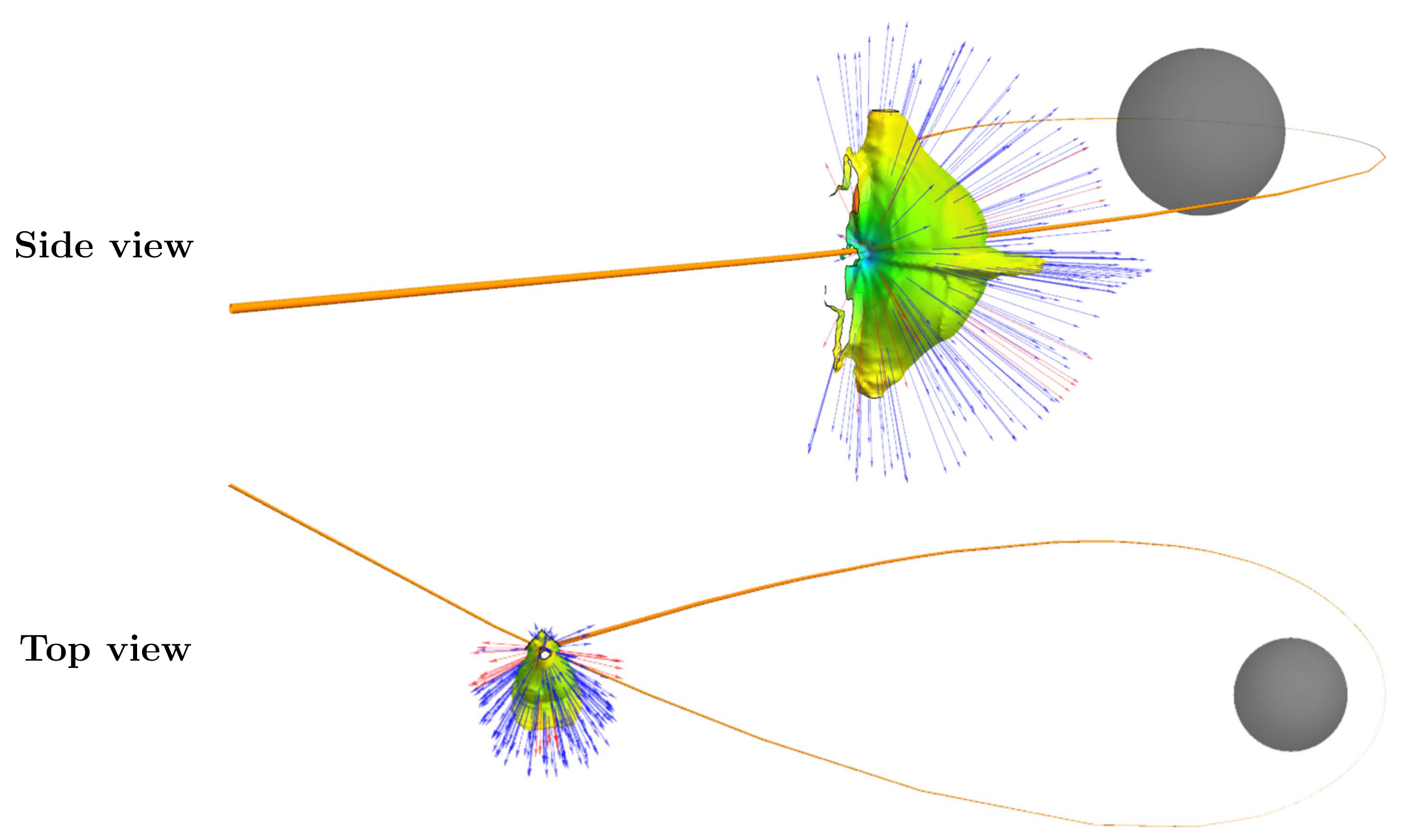} 
   \includegraphics[width=0.45\textwidth]{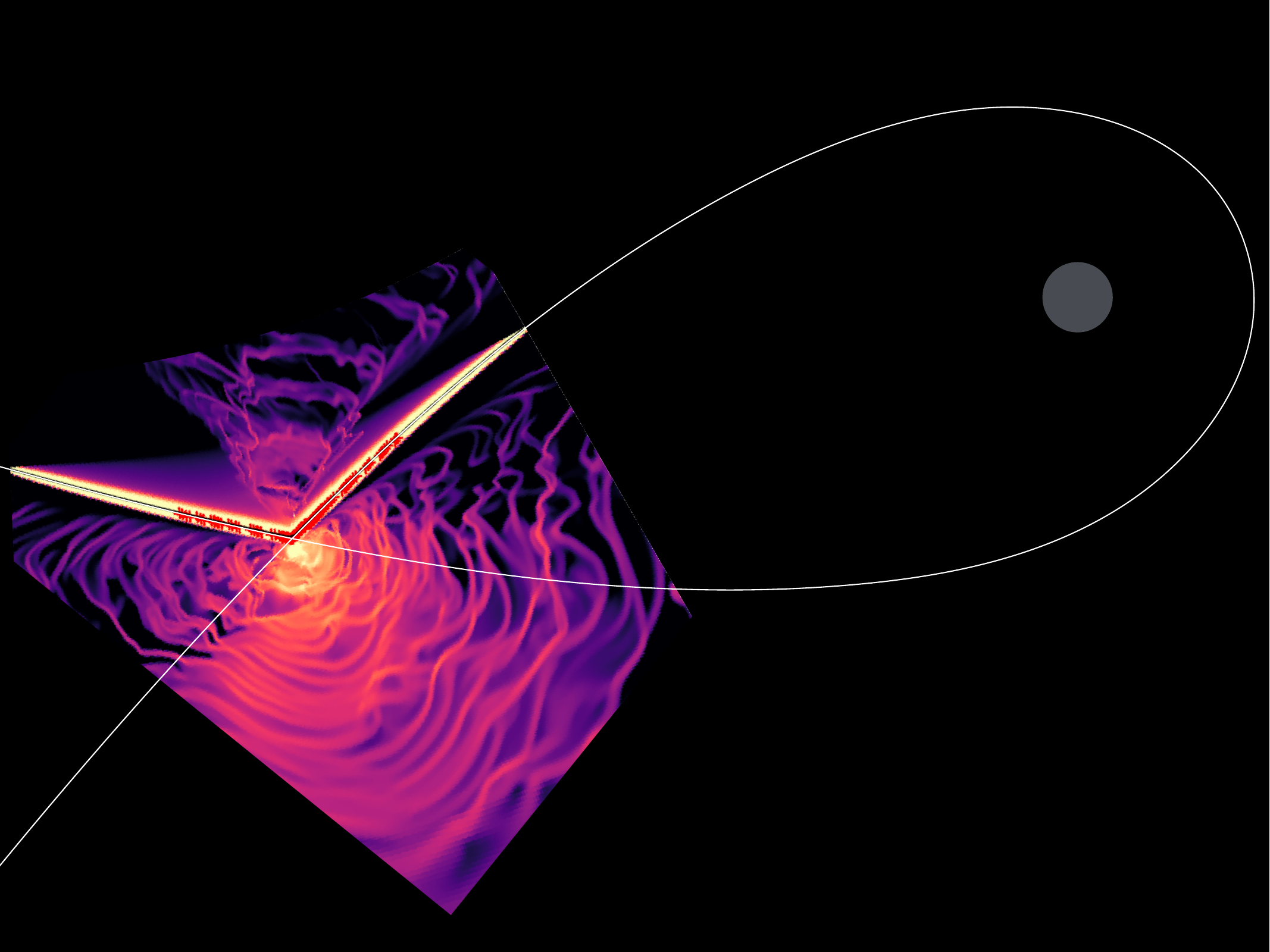} 
   \caption{Stream collision simulations superimposed on the preceding stream orbits. {\bf Left:} A snapshot from \citet{Jiang2016}, from a simulation with the following orbital parameters: velocity of collision $=0.3c$, collision angle $=135^\circ$, $\dot{M} = 0.33 \dot{M}_{\rm edd}$ (where $\dot{M}_{\rm edd}$ has been defined assuming an efficiency of $10\%$ and using the electron scattering opacity $\kappa_T$). This was found to match the following disruption parameters from the monte carlo stream collision parameter exploration in \citet{guillochon_dark_2015}: $M_\bullet = 10^{6.95} M_\odot$, spin $a = 0.035$, $\beta = 0.65$, $M_* = 0.2 M_\odot$ {\bf Right:} A snapshot from \citet{huang_bright_2023}, from a simulation based on the following disruption parameters: $M_\bullet = 10^{7} M_\odot$, $\beta = 1$ (for $M_* = 1 M_\odot$, $R_* = 1 R_\odot$), $a = 0$, and $\dot{M} = 1 \dot{M}_{\rm edd}$. In this case, the collision angle $=122^\circ$ and the velocity $= 0.17c$.}
   \label{fig:streamcollisions}
\end{figure} 

One type of these ‘circularizing’ interactions are stream collisions, where debris streams intersect and release energy (see Figure~\ref{fig:streamcollisions}). These collisions between adjacent orbits occur due to general relativistic apsidal precession  \citep{rees88}, and the stream intersection location can be calculated for a given initial orbit and star-SMBH system \citep[e.g.,][]{dai_soft_2015}. The more ‘head-on’ these collisions are, and the smaller the radii they occur at, the more energy will be dissipated by them \citep{kochanek_aftermath_1994, dai_soft_2015, jankovic_spin-induced_2023}. Stream collisions that dissipate significant energy often also lead to fast outflows that, while inherently aspherical, can still quickly enshroud the black hole \citep[e.g.,][]{Jiang2016, bonnerot_first_2021, huang_bright_2023}; we discuss evidence for outflows from TDEs in section \ref{sec: reprocess-spec}. If the black hole is spinning and there is significant out of plane precession, debris streams can miss each other on the initial orbit and delay this process \citep[e.g.][]{guillochon_dark_2015, batra_general_2023, jankovic_spin-induced_2023}. The importance of stream intersections in circularizing the gas is also dependent on the evolution of the stream’s density and cross-section – if the stream expands with time, or after nozzle shocks (described in more detail below), this will change its density and cross-section. The stream’s cross-section is dependent on its self-gravity, the effect of nozzle shocks on the stream, and the stream’s ability to cool, topics of current debate on which there is not yet a clear consensus \citep[e.g. ][]{bonnerot_nozzle_2022, steinberg_origins_2022, coughlin23}. 

It is also possible that nozzle shocks at pericenter may play a role in circularizing debris. Nozzle shocks occur when gas passes through pericenter, where the orbits of different parts of the gas stream intersect. Imagine gas orbiting in adjacent orbital planes with slight vertical offsets. At pericenter, these planes intersect, and in the frame of reference of the gas particles this looks like a vertical collapse, which produces a shock (similar to the ``pancake" shock at high $\beta$ described in Section~\ref{sec:highly_destructive}). While these shocks occur at pericenter, they are happening perpendicular to most of the stream’s velocity, and therefore do not have access to the majority of the KE in the stream (think of a car on a highway that gets into a grazing collision with a car moving parallel to it at a similar speed – the cars will bounce away from each other and likely suffer some damage, but the damage will be minimal compared to what you would get from a head on collision). Whether or not the nozzle shock releases sufficient energy to change the stream dynamics has been a topic of debate. Some work suggests stream dynamics are not dramatically affected \citep[e.g.][]{guillochon_ps1-10jh_2014, bonnerot_nozzle_2022}, and stream collisions at larger radii do the majority of the work circularizing debris. However, one paper \citep{steinberg_origins_2022} that modeled the entire disruption through to initial disk formation found that nozzle shocks dominated their circularization process. Their simulations had puffier streams from previous work, and the increased vertical extent led to higher vertical collapse velocities and stronger shocks\footnote{The reason the streams puff up in their simulation is recombination heating of debris, emphasizing the importance of accurately treating heating and cooling processes in the streams}. Other work has found that the stream width, the efficiency of heating and cooling of the streams as well as the nozzle shock strength can be very resolution dependent \citep[e.g.][]{bonnerot_nozzle_2022, coughlin22b, price_eddington_2024}, and at this point nozzle shocks are not resolved in simulations. It is also possible that different types of shocks are more important for different systems. Regardless, more work needs to be done to better understand the details of the circularization process. 

\begin{figure}[htbp] 
   \centering
\includegraphics[width=0.75\textwidth]{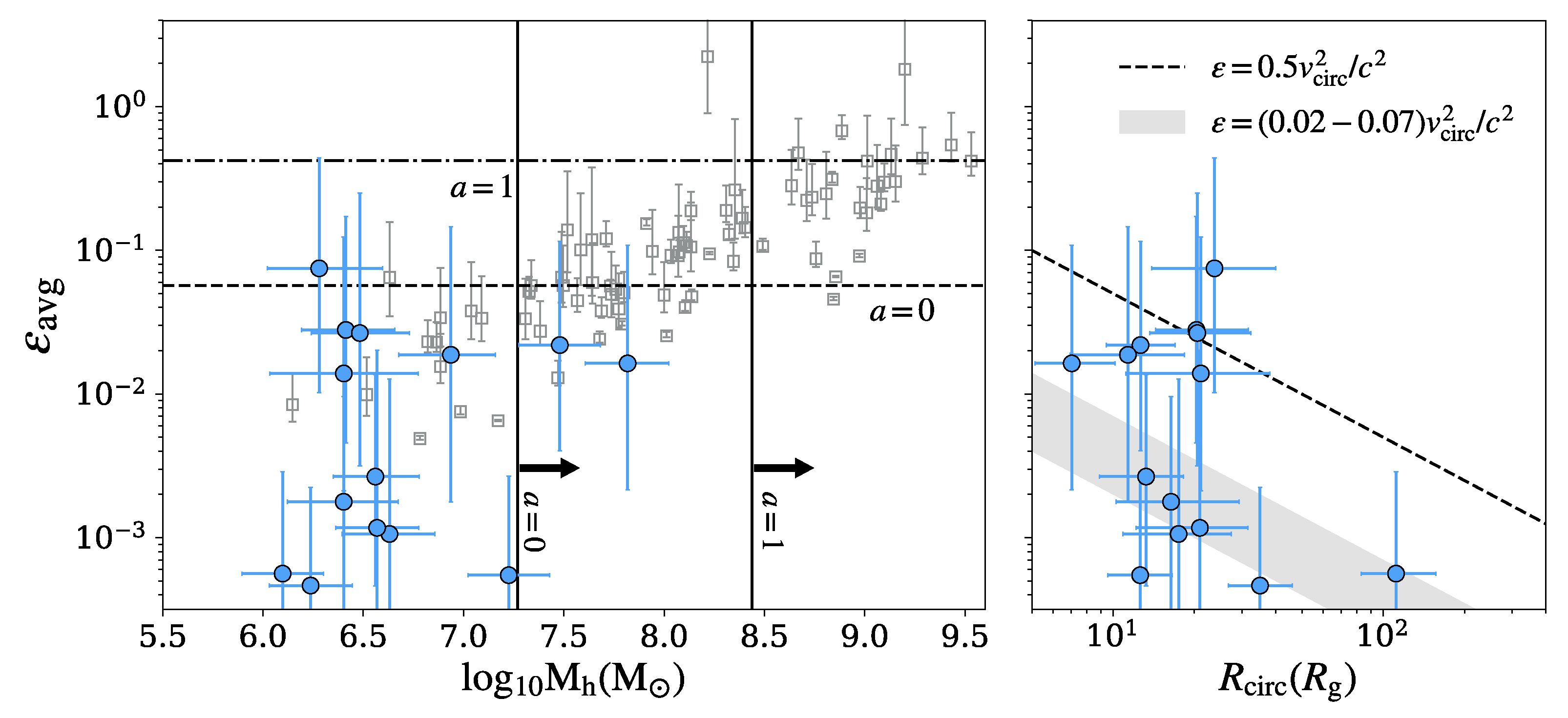} 
   \caption{Average efficiencies ($\int{L_{bb} dt}/\int{\dot{M} dt}$) estimated for the initial optical and UV flares of an early population of TDEs. 
   The large uncertainties in the efficiencies are mostly due to the degeneracy between the estimated efficiency and accreted stellar mass. {\bf Left:} Compared to observed AGN efficiencies from \citet{davis_radiative_2011}. The horizontal dashed and dash-dotted lines represent the estimated efficiencies for accreting black holes with spins of $a = 0$ and $a=1$ respectively. The vertical lines with arrows denote where $R_t < R_{\rm isco}$ for a $1 M_\odot$ zero-age main sequence star around a black hole again with spin of $a = 0$ and $a=1$ respectively. {\bf Right:} Compared to the estimated efficiencies for stream collisions at the circularization radius (most stream collisions will occur at larger radii, so this can be thought of as an upper limit on the stream collision efficiency). The dashed line assumes gas is perfectly virialized at the circularization radius, the gray shaded region uses more realistic stream collision efficiencies estimated from \citet[][]{Jiang2016}. The circularization radius is estimated from light curve fits with {\tt MOSFiT}, and is dependent on the mass of the black hole and the mass and radius of the star. However, given that for a given light curve increasing the mass of the star decreases both the estimated efficiency (because if $L$ is constant and $\dot{M}$ increases, $\varepsilon$ must decrease) and the maximum efficiency at the circularization radius (because $R_{\rm circ}/R_g$ increases with $M_*$), changing the stellar mass does not have a large effect on the ratio of the estimated efficiency to the efficiency of stream collisions at the estimated circularization radius. Adapted from \citet[][]{mockler_energy_2021}.}
   \label{fig:TDEefficiencies}
\end{figure} 

Eventually gas circularizes and forms a disk (see Figure~\ref{fig:dai}). As described above in Section~\ref{sec:howislum}, if a disk is formed near peak this will likely translate to super-Eddington accretion rates near the circularization radius. In this scenario, the type of disk formed will be quite different from a classic Shakura-Sunyaev thin disk. At accretion rates near Eddington, photons are trapped in the disk and advected inwards (reducing the maximum efficiency, though still much more efficient than the initial circularizing interactions), and the disk also puffs up due to radiation pressure, reaching H/R$\sim 0.3$ \citep{abramowicz_slim_1988, sadowski_slim_2009}. More recent work has also shown that in these somewhat confusingly named `slim' disks, the radiation pressure can drive winds that surround the black hole out to larger radii \citep{mckinney_three-dimensional_2014}. For TDE-like disk parameters, this can produce a photosphere that looks quasi-spherical at most viewing angles \citep[with the exception of viewing angles near the poles,][see Section~\ref{sec: reprocess-spec}]{Dai2018, thomsen_dynamical_2022}. Once the accretion rate decreases and the disk luminosity drops to some fraction of Eddington \citep[ $\lesssim 0.3 L_{\rm edd}$, e.g. ][]{abramowicz_foundations_2013}, the disk would be expected to settle into a more efficient classic `thin' disk. 

\subsection{What sets the emission timescale?}

\begin{figure}[htbp] 
   \centering
   \includegraphics[width=0.5\textwidth]{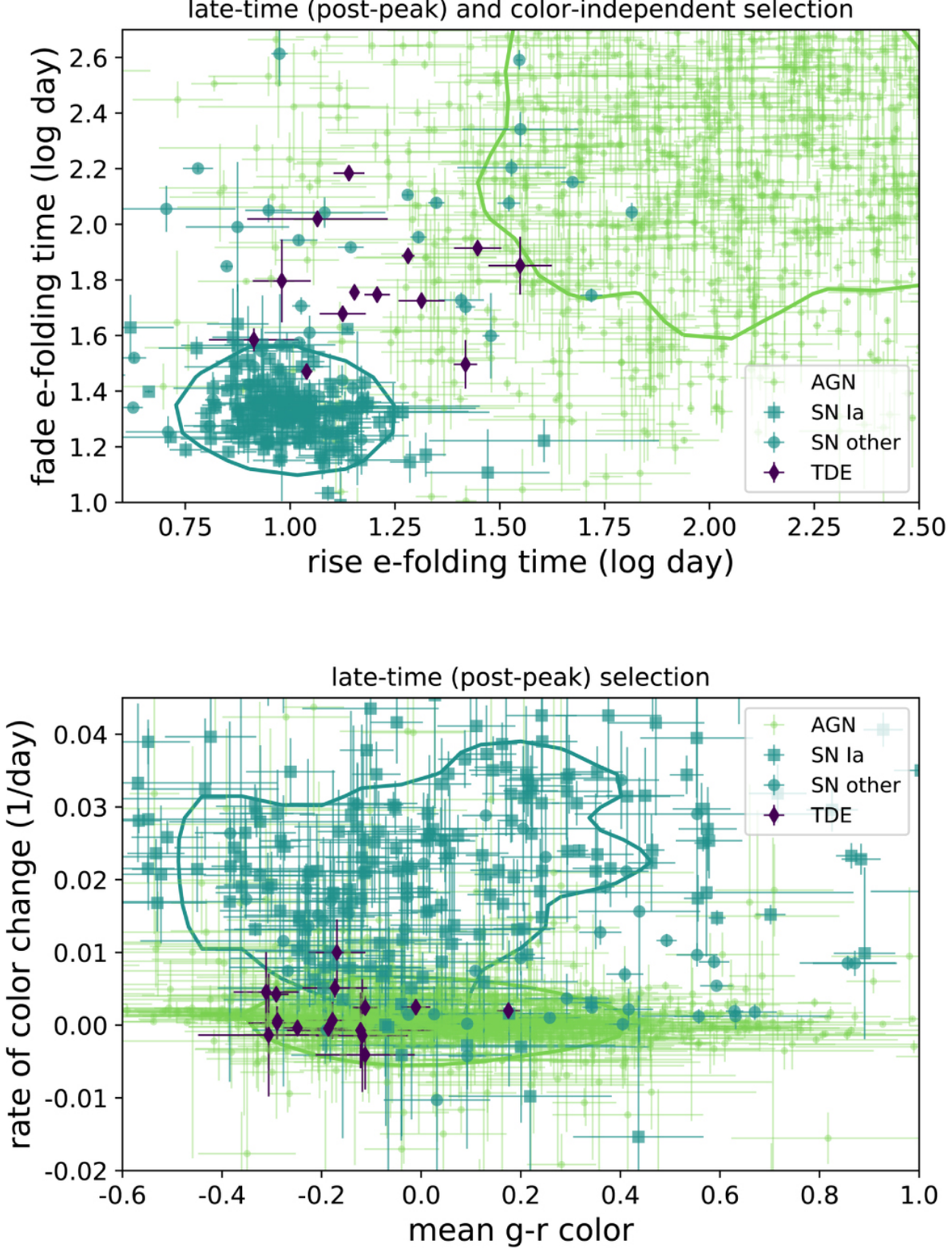}
   \caption{Rise/decline timescales and color evolution for the first 17 TDEs discovered by ZTF, compared against supernovae (teal) and AGN (green). For these background populations, contours containing two-thirds of each type of transient are overplotted to guide the eye. {\bf Top:} TDEs evolve on longer timescales than most supernovae (in teal) and shorter timescales than most AGN (in green). {\bf Bottom:} TDEs are blue (small or negative values of g-r) and their color is relatively constant with time compared to supernovae. This behavior is much more similar to other accreting black holes, e.g. AGN. Figure from \citet{vanVelzen2021}.}
   \label{fig:emissiontimescale}
\end{figure}

\begin{figure}[htbp] 
   \centering
\includegraphics[width=0.75\textwidth]{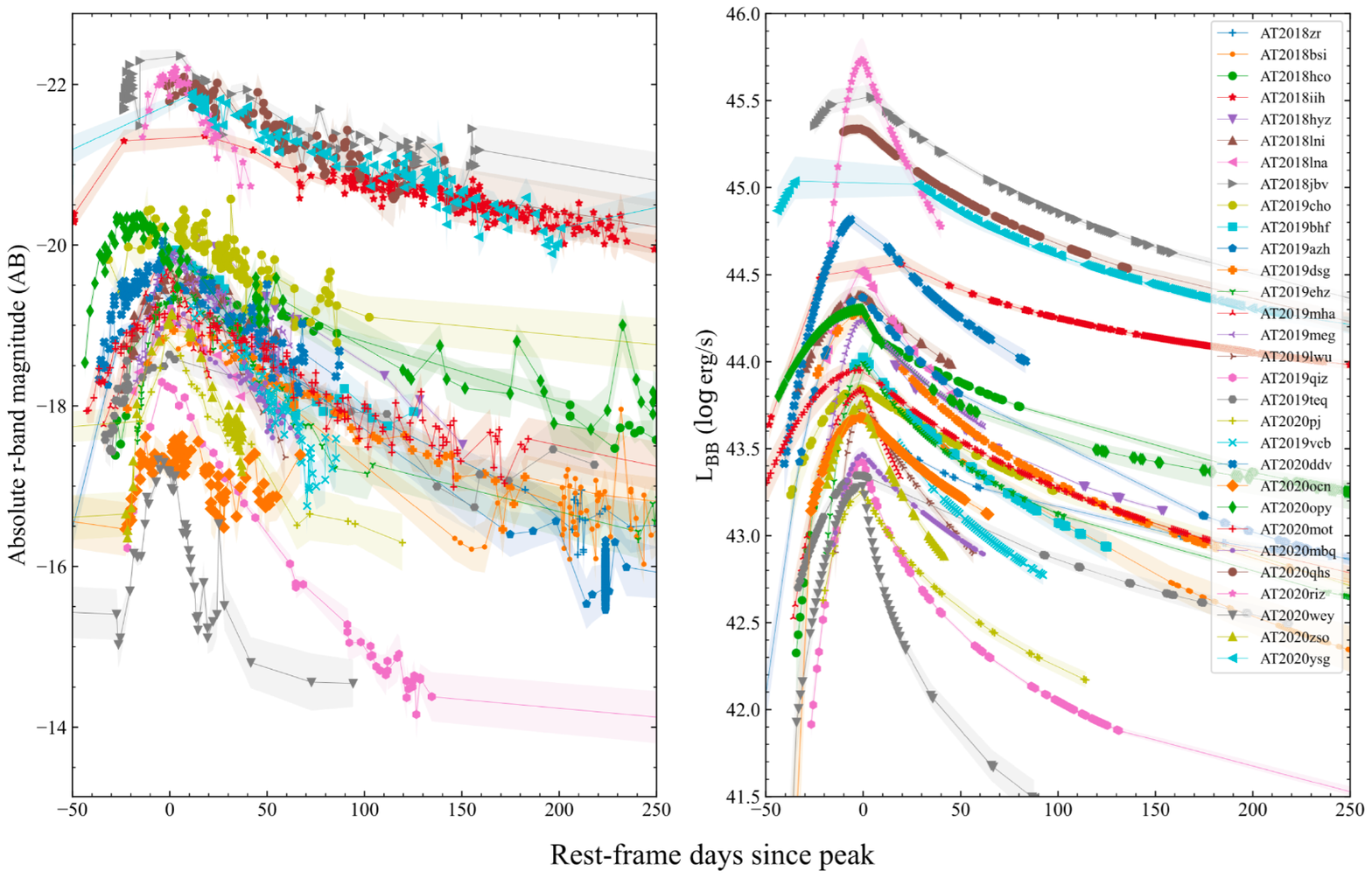} 
   \caption{{\bf Left:} Observations of TDE light curves in R-band. The shaded regions denote the size of the errorbars. Note that the true error is likely larger, and the small timescale variability may not be real, as variability increases near the detection limit. {\bf Right:} The blackbody luminosity curves obtained by fitting the optical and UV data for the same events. To obtain the blackbody luminosity, temperature was fit periodically throughout the light curve at points where UV data was available (on 30 day intervals), and was required to be smoothly varying in between these points. Figure from \citet{Hammerstein2023}.}
   \label{fig:lightcurves}
\end{figure}

TDEs evolve on similar timescales to supernovae, with rise times of $\sim$ weeks to months, however on average they are somewhat longer-lived. For example, in \citet{vanVelzen2021}, the authors show the e-folding decay times of TDEs in their sample are longer than almost all nuclear type Ia supernovae, and more generally than the majority of nuclear supernovae of all types found in the Zwicky Transient Facility (ZTF) transient survey (see Figures~\ref{fig:emissiontimescale} and \ref{fig:lightcurves}). This is consistent with theoretical expectations, as the luminosity source of TDEs is also generally longer-lived than the luminosity sources in supernovae. Theoretically, a TDE’s luminosity is ultimately limited by the mass fallback rate to the black hole. There are additional processes that can act to prolong the luminosity timescale compared to the mass fallback timescale, such as diffusion through a dense envelope and inefficient circularization, but it is difficult to shorten it. After peak, the mass fallback rate of a full disruption approaches $\propto  t^{- 5/3}$, decaying more slowly than the exponential heating rate from radioactive decay in supernovae \citep[the e-folding timescale of Ni56 = 8.8 days, e.g.][]{branch_supernova_2017}, and even slightly more slowly than the dipole spin down of a magnetic neutron star ($\propto t^{-2}$), thought to potentially power some longer-lived supernovae \citep[e.g.][]{kasen_supernova_2010, woosley_bright_2010}. The mass fallback rate of a partial disruption evolves a bit faster -- approximately $\propto t^{-9/4}$ \citep{coughlin19}, though still slower than exponential decay. The disk viscous timescale will only dominate over the mass fallback rate if its viscous timescale is long enough that mass builds up in the disk, in which case it will act to further extend the light curve.

Observations of many TDE light curves near peak appear to follow rises and power law declines similar to those predicted by theoretical mass fallback rates. This correlation between the shape of the light curve and the mass fallback rate is predicted by some models of emission originating from stream collisions and shocks whose luminosity is roughly dependent on the mass inflow rate \citep{Jiang2016, steinberg_origins_2022, ryu_shocks_2023, huang_bright_2023}, and also by models with the emission originating from a disk whose viscous accretion timescale is shorter than the fallback timescale of debris \citep[e.g.][]{rees88, Dai2018, bonnerot_first_2021}. In \citet{vanVelzen2021} and \citet{Hammerstein2023}, the authors fit estimates of the post-peak bolometric luminosities of the population of ZTF TDEs with power laws of the form of $t^{-\chi}$. While there is spread in their results, the distributions of their best fits for $\chi$ peaked between 1.4 and 1.9, consistent with the theoretically predicted $\chi$ values for full and partial disruptions ($\chi = 5/3$ and $\chi = 9/4$ respectively). Similarly, \citet{mockler_weighing_2019} and \citet{Nicholl2022} found that the light curves of most optical and UV TDEs can be well modeled with luminosities that approximately follow the predicted mass fallback rates from hydro simulations. This is compelling, because if we can constrain the timescale of the fallback rate from the luminosity, we can use it to constrain the black hole's mass. \citet{mockler_weighing_2019} and \citet{van_velzen_optical-ultraviolet_2020} found that the timescales of TDE light curves scale with the black hole mass as predicted theoretically for the mass fallback rate ($t \propto M_\bullet^{1/2}$). \citet{Yao2023} also found that the timescale between half maximum light before and after peak increases with black hole mass, although they found a weaker relation than predicted theoretically ($t_{1/2} \propto M_{\bullet}^{0.17 +- 0.15}$). 

However, models that do not rely on the luminosity following the fallback rate can sometimes produce similar behavior. For example, in \citet{sarin_tidal_2024}, the authors find that a model of a TDE as a cooling Eddington envelope can also fit many of the observed light curves. In this scenario, energy emitted from the process of circularizing debris is trapped in a large, optically thick envelope and slowly leaks out as the envelope cools \citep[similar to the model proposed in][see Section~\ref{sec: reprocess-spec} for more details]{Loeb1997}. The observed light curve timescale is then dependent on the cooling timescale, not on the timescale of the source emission, whether it be from shocks or accretion processes. 

\subsection{Early and late time light curves}\label{sec:earlylatetime}

The first light curves of TDEs were only observed near peak. However, over the past $\sim 5$ years, transient surveys have found TDEs at earlier times on their rise and also continued to follow up known candidates for years after peak. Data at these very early and late times often show divergences from the time evolution predicted by theoretical fallback rates (a fast rise and power law decline). For example, many TDEs observed at late times have shown signs of plateaus in their luminosity decay years after their initial discoveries \citep[e.g.][]{van_velzen_late-time_2019, Hammerstein2023, Yao2023}. This has been explained by a viscously spreading disk beginning to dominate the light curves \citep{van_velzen_late-time_2019, mummery_fundamental_2023}. Constraining the properties of such a disk can provide another avenue for estimating the black hole mass, independently of the fallback timescale \citep{mummery_fundamental_2023}. For example, it has been shown theoretically that the late-time luminosity at UV and optical wavelengths should be approximately constant for a given system\footnote{Importantly, this is not because the total luminosity is constant. Rather, it is because as the luminosity decreases, so does the emitting area of the disk, leading to a larger fraction of the total luminosity being emitted at optical and UV wavelengths at later times. This balances out the effect of the decreasing total luminosity at these wavelengths.} and increase with black hole mass as $L \propto M_\bullet^{2/3}$. Observations have found that the late-time plateau luminosities do increase with black hole mass as predicted theoretically.  Most of these events have relatively sparse data, making it difficult to constrain the shape of the SED \citep[e.g.][]{van_velzen_late-time_2019}. However, if the emission mechanism is an unobscured disk, it is more straightforward to model than earlier emission mechanisms. 

Even more recently, a few TDE light curves have shown evidence for variability at very early times that could potentially be explained by circularization shocks \citep[such as][]{faris_light-curve_2023, huang_AT2023lli_2024}. The vast majority of observed TDE light curves only have pre-peak data within one magnitude of peak (if they have any pre-peak data at all), and so it is only recently and for a small subset of events that we are able to analyze the early part of the rise.

\subsection{Constraining the total emitted energy}

\begin{figure}[htbp] 
   \centering
   \includegraphics[width=0.42\textwidth]{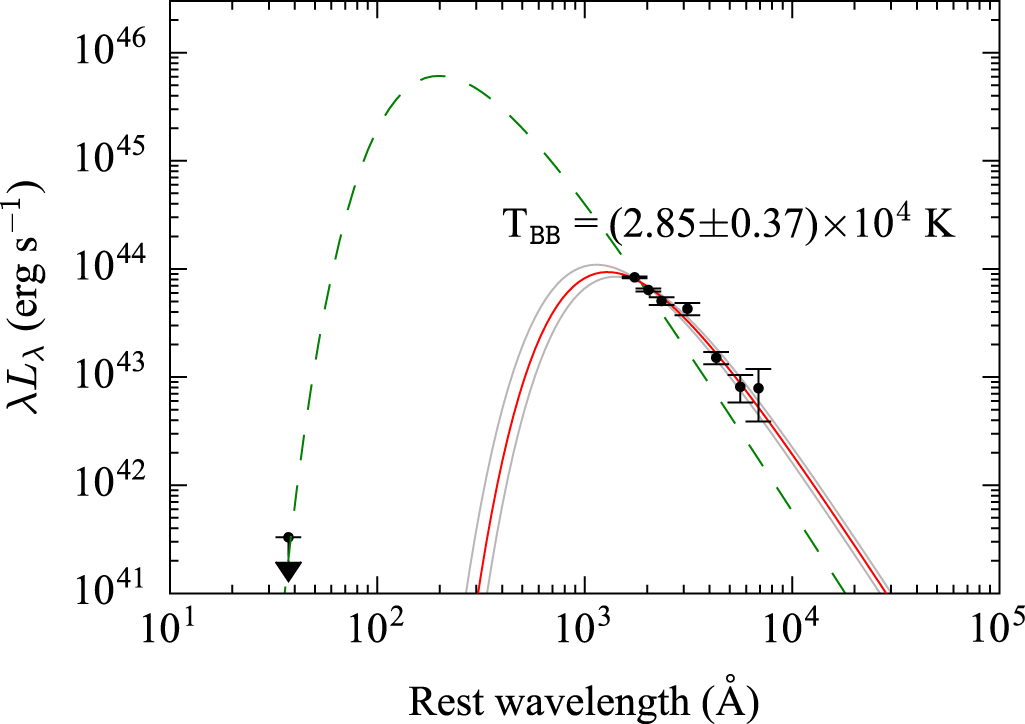} 
   \includegraphics[width=0.53\textwidth]{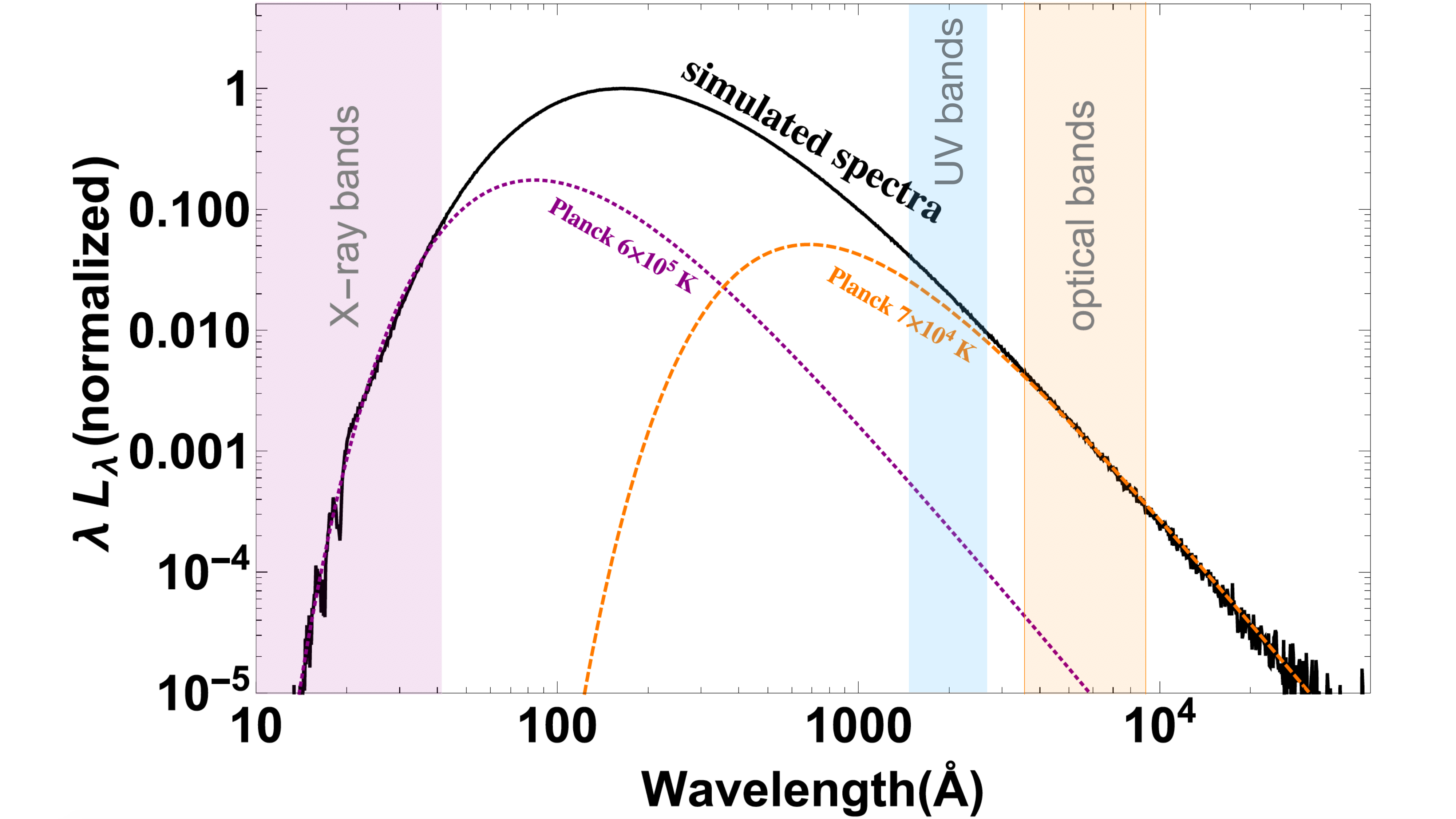} 
   \caption{{\bf Left: } For this example SED, the UV photometry and upper limits in the X-ray put some constraints on the peak temperature of the SED, and the data is reasonably well fit by a single blackbody. However, for events like this, small differences in UV luminosities can result in very different best fit blackbody curves because the peak of the SED is not measured directly. For example, different assumptions for the extinction (which will change the estimated UV luminosity) can easily result in under- or over-predicting the total energy released in the flare. {\bf Right: } The simulated spectra in black is from a post-processed radiation transport calculation of what spectra would look like from a super-Eddington accretion disk simulation with TDE-like properties. In this case, the spectra is well-approximated by the sum of two blackbodies peaking at different temperatures. However, if it were only observed at optical and UV wavelengths (the plotted UV wavelength range corresponds to the bandpass for the Swift telescope), it would be easy to underestimate the true luminosity by fitting the SED with a single blackbody that peaked at lower temperatures. {\bf Left: } Figure from \citet{hung_revisiting_2017} for the TDE iPTF16axa. {\bf Right: }Figure from \citet{Dai2018}. }
   \label{fig:SEDs}
\end{figure}

Estimating the total emitted energy in a TDE (or any astrophysical phenomenon) requires some model for the SED as a function of time. For a TDE, this often requires either making extrapolations or taking lower limits. This is because most optical and UV TDEs appear to peak in the UV, but observations of the UV spectrum are not available for most TDEs, and many lack any photometric data blueward of the peak of the SED (see Figure~\ref{fig:SEDs} and \citealt{hung_revisiting_2017} for a detailed example). When we fit the photometry with a blackbody model, this means we must try to constrain the initial decline of the SED before it approaches its power-law tail. We therefore rely strongly on the UV data, and degenerate fits are common (see Figure~\ref{fig:SEDs}).

These sorts of fits to observations of TDEs find that the total energy emitted in the initial optical and UV flares in these events (not including late time plateaus) is generally between $10^{50} - 10^{52}$ ergs \citep[e.g.,][and individual event citations therein]{mockler_energy_2021}. This is less than the originally predicted $\sim 10^{53}$ ergs from multiplying $0.5 M_\odot$ by $0.1 \times c^2$ \citep[where $0.1$ is the canonical thin disk efficiency seen for many AGN, e.g.][]{davis_radiative_2011, abramowicz_foundations_2013}. This discrepancy has been termed the ``missing energy problem". However, there are many caveats to this back of the envelope calculation that can lower the total observed luminosity. 

First of all, if we expect the masses of stars disrupted in TDEs to mostly track the masses of stars in the nuclear cluster, we would expect most disruptions to be of stars $\lesssim 0.5 M_\odot$, as lower mass stars dominate even if the initial mass function is top-heavy \citep[such as that estimated for our own galactic center, e.g.][]{hosek_unusual_2019}. Second, not all of the mass returns at once. Only about half of the stellar debris will return near peak, further reducing the mass budget. Additionally, the TDE system, especially at early times, is very different from a thin disk AGN system. If it has formed an accretion disk, it will be super-Eddington, and simulations show the accretion efficiency will be significantly reduced \citep[super-Eddington disk simulations find mass to energy efficiencies of $0.01-0.05$, e.g.][though this is dependent on how super-Eddington the accretion rate is]{abramowicz_foundations_2013, mckinney_three-dimensional_2014, dai_soft_2015}. If an accretion disk hasn't formed and luminosity is coming solely from shocks, the efficiencies will be even lower \citep[ $\lesssim 0.1 \times R_{\rm collision}/R_{\rm g}$, e.g.][]{Jiang2016, huang_bright_2023}. 
Figure~\ref{fig:TDEefficiencies} looks at the efficiencies estimated for the initial optical and UV flares from an early subset of TDE light curves and compares them to efficiencies estimated for AGN (from observations) and stream collision efficiencies (from simulations). The efficiencies estimated for the TDEs are generally consistent with emission on the size scale of a forming disk, and are mostly higher than would be expected for stream collisions at larger radii. However, there are large uncertainties in the efficiency estimates, as they are somewhat model dependent, and this is also a small subset of the current TDE candidate population. If material is also lost to collisionally induced outflows (from stream collisions) or radiation driven winds (if the mass accretion rate is near the Eddington limit), this will further reduce the overall efficiency in producing luminosity from the mass falling back. Finally, these blackbody estimates are necessarily lower limits to the total observed luminosity. Additional energy could be missed at EUV wavelengths, as shown in Figure~\ref{fig:SEDs} and described in \citet{lu_missing_2018, thomsen_dynamical_2022}, and in late-time plateaus in the emission, as described in Section~\ref{sec:earlylatetime}.   

In a few cases, IR dust emission has been used as a bolometer to constrain the total amount of energy released in an optical/UV bright TDE long after the initial flare (\citealt{vanVelzen2016, Jiang2021, dou_long_2016}, review by \citealt{van_velzen_optical-ultraviolet_2020}). Dust can absorb emission at far UV wavelengths near the predicted peak of the SED but outside of the bandpass observed in the initial observations, and then re-emit this energy at IR wavelengths. These works find higher values for the total bolometric energy released than was found using blackbody models of the initial observed light curves, emphasizing that our fits to the initial flares often underestimate the total amount of energy released by these transients. More recently, a population of IR selected TDE candidates has also been found, and this population similarly finds higher overall integrated energies than events only observed at optical and UV wavelengths \citep{Masterson2024}. 

\section{The observed temperatures and spectra of TDEs} \label{sec: reprocess-spec}

\subsection{X-ray vs optical emission}

We have seen above that a TDE appears observationally as an increase in luminosity from the nucleus of a galaxy, followed by a slower power-law decay. The first sources matching this description were discovered in the 1990s at X-ray wavelengths \citep{Bade1996,Grupe1999,Komossa1999}. The temperature of the detected emission from these early candidates was $\sim10^5-10^6$\,K. TDEs stand out from common AGN in X-ray surveys because of their thermal spectra, which are `soft' (peak at lower energies) compared to the `hard' power-law spectra of AGN (thought to originate from inverse Compton scattering in a `corona' of ionized gas). From the Stefan-Boltzmann law, radiation at this temperature with a luminosity of $\sim 10^{44}$\,erg\,s$^{-1}$ (the Eddington luminosity of a $10^6$\,M$_\odot$ BH) would correspond to an emitting region with a characteristic size of $\sim10^{11}-10^{12}$\,cm, comparable to the innermost stable circular orbit. This was consistent with some theoretical predictions for emission from accretion disks in TDEs \citep{Cannizzo1990,Ulmer1999}.

However, with UV detections in the 2000s \citep{Gezari2006,Gezari2008} and the optical TDE discoveries that have dominated the detection rate since 2010 \citep[beginning with][]{vanVelzen2011,Gezari2012,Holoien2014}, a very different picture has emerged. Most of these candidates, often referred to as `optical TDEs', show much lower blackbody temperatures of $\sim {\rm few}\times 10^4$\,K, giving characteristic sizes hundreds of times the disk scale. Although their optical luminosities evolve on timescales roughly comparable to some supernovae, TDE temperatures are higher and in many cases do not change substantially over time (unlike supernovae, which cool adiabatically). The combination of blue colors, which can be measured quickly from broadband photometry in multiple filters, and lack of color change over time, can be used to identify candidate TDEs in all-sky imaging surveys \citep[][Figure \ref{fig:lightcurves}]{Hung2018,Bricman2023}.

The inferred radiation temperatures of the first $\sim 50$ known TDEs were compiled by \citet{Gezari2021}, highlighting a bimodality between events discovered in X-ray surveys, exhibiting $T\sim10^6$\,K, and those found in optical surveys, exhibiting $T\sim10^4$\,K. A substantial fraction ($\gtrsim 40\%$) of TDEs discovered in the optical have been subsequently detected in X-rays via targeted follow-up observations \citep{Guolo2023}. The X-ray to optical ratio in these sources spans orders of magnitude, from $\sim 1$ down to $\lesssim10^{-3}$ \citep{Auchettl2017,Guolo2023}. Most of the optically-detected events with X-rays show characteristic temperatures $\sim10^6$\,K in their X-ray spectra, inconsistent with their optical spectra but comparable to the TDEs discovered in X-ray surveys. The implication is that the initial X-ray and optical emission must arise from separate spatial regions, even when both are detected in the same event. After a few hundred days (during the plateau phase of the light curve; see previous section), the accretion disk is more easily visible, and may produce both the X-ray and UV/optical emission \citep{van_velzen_late-time_2019,Jonker2020,Wen2020,Mummery2023,Mummery2024,Guolo2023}.

\subsection{What produces the optical colors in TDEs?}

While X-ray emission can naturally be associated with the innermost (hottest) region of an accretion disk, the origin of the cooler optical emission has been heavily debated. The simplest proposal is that high-energy photons from the disk do not always reach the observer directly, but are absorbed by an extended atmosphere and re-radiated at an outer photosphere with a correspondingly lower blackbody temperature. The earliest such reprocessing model, from \citet{Loeb1997}, actually predates the optical discovery of TDEs. In this model, accretion at the Eddington rate slows the infall of debris due to radiation pressure, forming a quasi-spherical `envelope' in hydrostatic equilibrium if the debris cooling time is longer than its dynamical time. This model is analogous to stellar structure, but with accretion rather than fusion as the central radiation source. The predicted scale of $\sim 10^{15}$\,cm for this `Eddington envelope' is comparable to the observed blackbody radii in the brightest TDEs. The `cooling envelope' model proposed by \citet{Metzger2022} is conceptually similar, but additionally postulates that the main source of radiation comes from the envelope itself, which must lose energy in order to contract to the scale of the exposed disks seen at later phases (Section \ref{sec:luminosity_production}).

\begin{figure}[t]
\centering
\includegraphics[width=.6\textwidth]{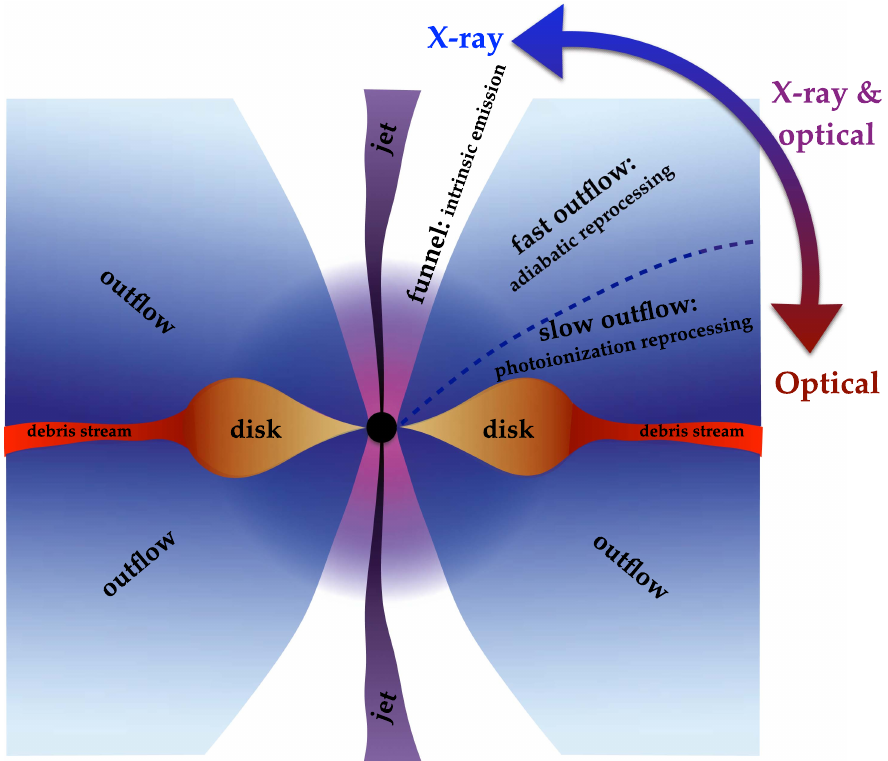}
\caption{Angle-dependent envelope model from the simulations of \citet{Dai2018}. The reprocessing layer is composed of outflows driven by radiation pressure in the accretion disk. A polar cavity allows X-rays to escape to some observers. From \citet{Dai2018}.}
\label{fig:dai}
\end{figure}

Other reprocessing models are even more dynamic, and analogous to supernovae rather than stars. We have seen that super-Eddington accretion rates are expected soon after disruption. Winds driven by radiation pressure can trap photons and emit them at larger radii as the photosphere expands. \citet{Strubbe2009} predicted that super-Eddington outflows with characteristic temperatures of $\sim{\rm few}\times10^4$\,K would dominate optical TDE detection rates. However, their model also predicted that TDEs would reach significantly higher temperatures within a few days, radiating mainly in the extreme UV, as hotter regions are exposed by outflows with $v\sim c$. An alternative wind model was developed by \citet{Metzger2016}, who assumed that only a small fraction $f\ll1$ of gas would be accreted during the super-Eddington phase. This leads to winds that contain most of the disrupted stellar mass, with relatively modest velocities $\sim10^4$\,km\,s$^{-1}$. By remaining optically thick for weeks, this can produce the observed colors (and slow color evolution) of observed TDEs, as hard photons from the inner accretion flow are degraded to lower energies by diffusion and advection in the dense wind.

None of these models easily accounts for the wide diversity in observed X-ray luminosity. Due to the very different size scales of the X-ray and optical emitting regions, an appealing possibility is a viewing angle dependence: if the reprocessing layer is non-spherical and so does not fully surround the X-ray source, a TDE can be X-ray bright when observed along favorable lines of sight. Simulations of the super-Eddington accretion phase by \citet{Dai2018} naturally produced a geometrically thick accretion disk that had both wide-angle winds with a range of velocities and a cavity directly above and below the disk plane (Figure \ref{fig:dai}; see also Figure \ref{fig:SEDs}). X-rays can escape through this cavity for observers close to the pole, whereas for observers closer to the disk plane, X-rays are absorbed in the wind and re-emitted at UV and optical wavelengths. In this case, the fraction of TDEs with no X-rays at early times is determined simply by the covering fraction of the wind.

As an alternative way to produce an envelope without super-Eddington accretion, simulations of the stream collision phase in TDEs have shown that the self-intersection shocks can also drive mass outflows \citep{Jiang2016}. Intuitively, if the streams collide close to the local escape speed (a condition met in disruptions with strong apsidal precession), a large fraction of the gas will become unbound. This has been termed a `collision-induced outflow' \citep{Lu2020}, and can also play the role of a reprocessing layer to produce optical TDEs. Launching these outflows off-center introduces an asymmetry, where the optical depth to X-rays may be lower for some viewing angles. This assumes accretion is still the ultimate energy source; if the stream collisions themselves are the source of luminosity, this will naturally produce optical rather than X-ray TDEs if the characteristic size of the emitting region is set by the intersection radius $\sim 10^{14}$\,cm \citep{Piran2015}. However, this may not easily explain why a separate X-ray component appears in many optical TDEs.

\subsection{Spectroscopic properties of TDEs }

\begin{figure}[t]
\centering
\includegraphics[width=.8\textwidth]{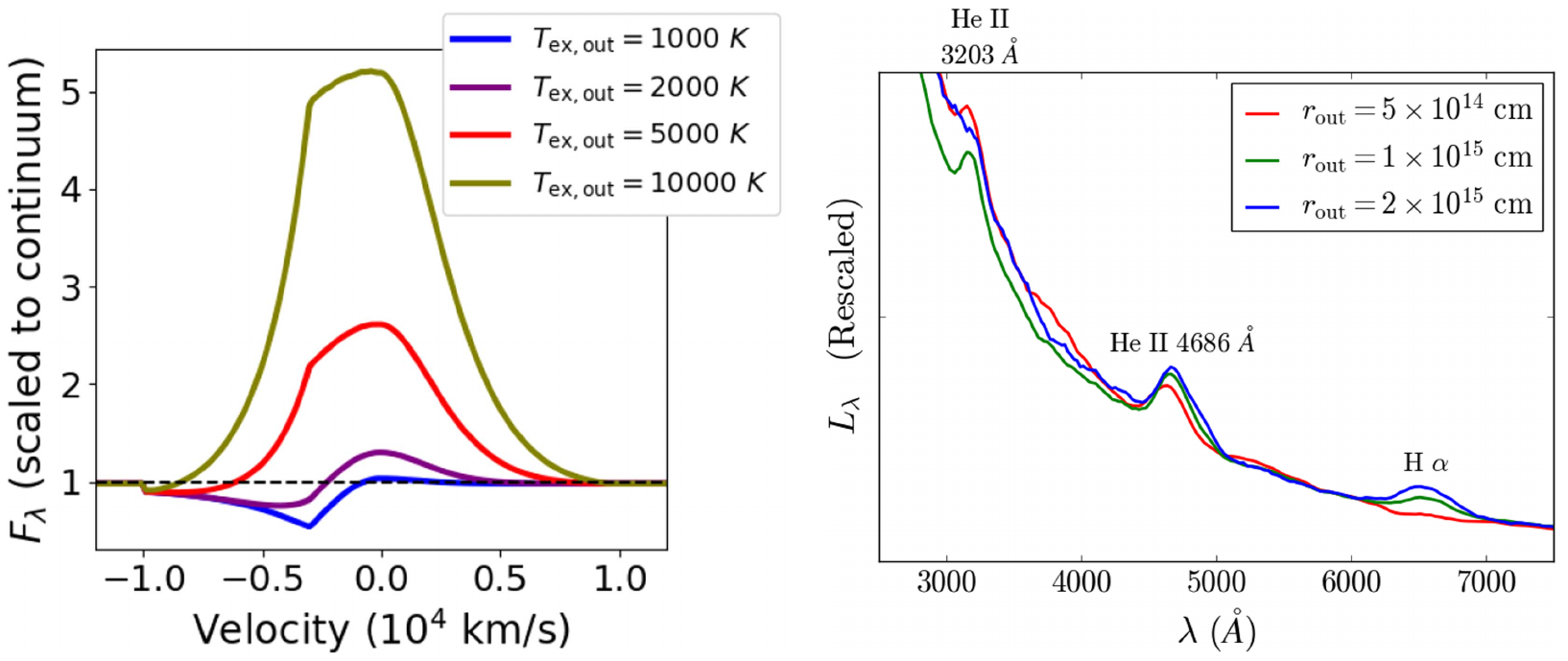}
\caption{Spectral line formation in TDEs. {\bf Left:} Line profile dependence on the `excitation temperature' (a measure of the electron population in excited states) in the emitting region. At low temperatures, a P Cygni profile arises due to absorption in gas moving towards the observer. In TDEs, the emission component is completely dominant due to the high temperature and additional non-thermal excitation by X-ray and UV photons. {\bf Right:} Sensitivity of line ratios to the size of the reprocessing layer. For more compact photospheres, H$\alpha$ is optically thick and becomes self-absorbed, changing the observed ratio of the hydrogen and helium line strengths. From \cite{Roth2016} and \cite{Roth2018}.}
\label{fig:roth}
\end{figure}

Spectroscopy of optical TDEs shows that in addition to the hot blackbody continuum, most also display broad atomic/ionic emission lines. These lines have velocity widths ($\Delta\lambda/\lambda=v/c$, where $\lambda$ is the rest wavelength of the line) of $v\sim{\rm few}\times 10^3-10^4$\,km\,s$^{-1}$. While the broad-line regions of AGN show comparable line widths, AGN lines (driven by photo-ionization and recombination in a large reservoir of gas) have much stronger and sharper peaks. In most cases, this enables spectroscopic separation of TDEs from other activity in the centers of galaxies, though this is more complicated if the galaxy also harbors a pre-existing AGN \citep{Blanchard2017,Cannizzaro2022,Petrushevska2023}. The term `Ambiguous Nuclear Transient' has been applied to events showing light curves similar to TDEs but spectra more comparable to AGN \citep[e.g.][]{Frederick2021,Holoien2022,Hinkle2022,Oates2024}. TDE spectra also look very different from supernovae. Supernovae show broad scattering peaks superimposed on blue-shifted absorption (a `P Cygni' line profile). Although their debris has similar mass and velocity, in a TDE this gas is subject to copious UV and X-ray flux, leading to additional non-thermal line excitation. The effective `excitation temperatures' are $\gtrsim10^4$\,K, producing an almost pure emission spectrum \citep[Figure \ref{fig:roth};][]{Roth2018}.

\begin{figure}[t]
\centering
\includegraphics[width=.6\textwidth]{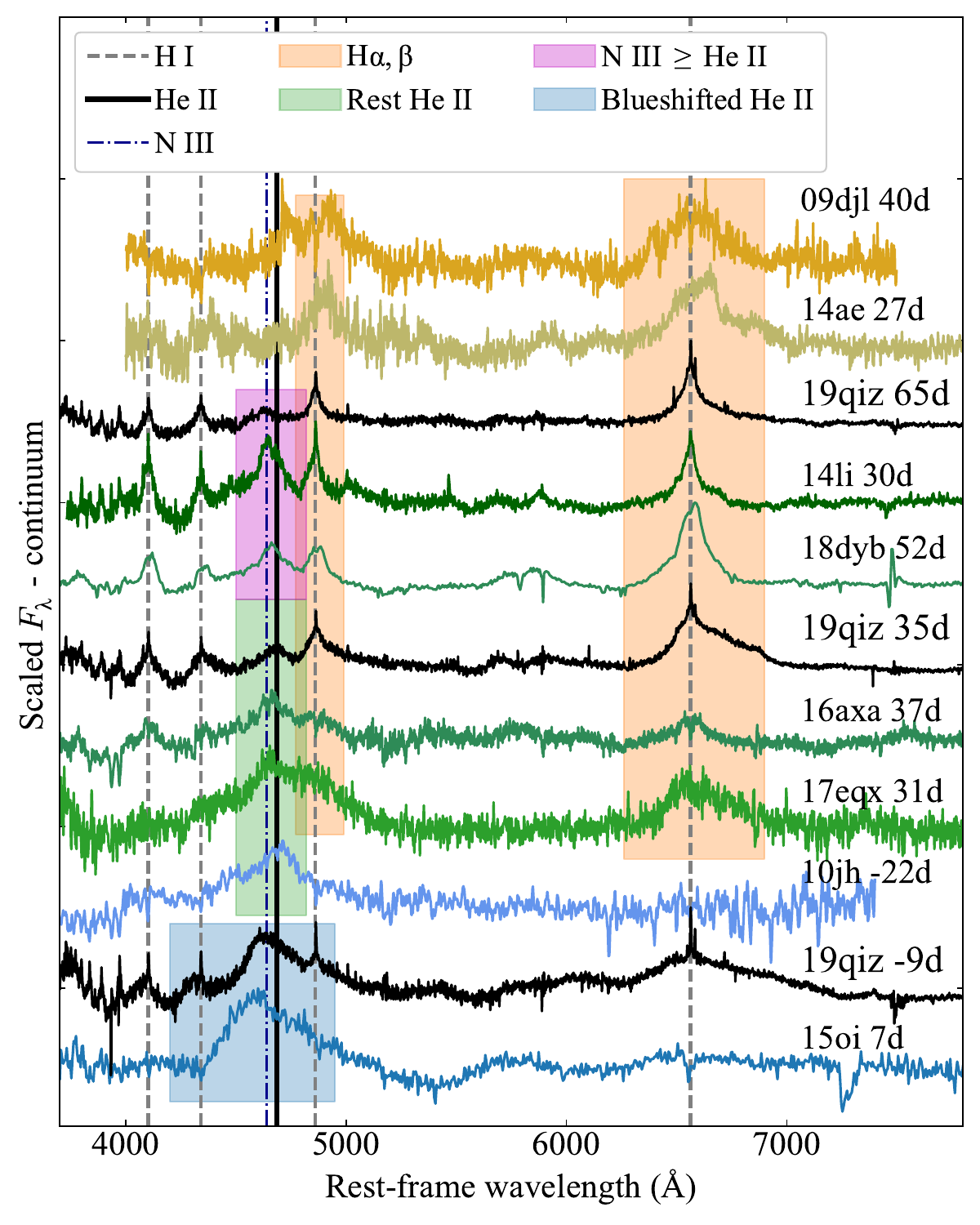}
\caption{Observed TDE spectra exhibiting a range of line profiles and ratios. The underlying blackbody continuum has been subtracted to emphasize the lines. From top to bottom, the TDEs shown range from TDE-H, through TDE-H+He, to pure TDE-He. Each spectrum is labeled with the name of the TDE, and the time when the spectrum was obtained relative to the time of maximum luminosity. Some TDEs show blueshifted lines from outflows, which are more easily discerned at early times when the outflows are optically thick. Figure from \citet{Nicholl2020}.}
\label{fig:spec}
\end{figure}

The electronic transitions producing the dominant lines around the time of the light curve peak vary between TDEs \citep{Arcavi2014}, leading \citet{vanVelzen2021} to coin a number of spectroscopic sub-classes. These are shown in Figure \ref{fig:spec}. Type TDE-H exhibit only hydrogen lines (the Balmer series at optical wavelengths). TDE-H+He also show Balmer lines, along with a strong line from singly-ionized helium (He II) at 4686\,\AA. About half of TDEs fall into this class, and most of those also show emission from nitrogen (N III) at 4640\,\AA. This line arises through the Bowen fluorescence mechanism \citep{Bowen1935}, a cascade of resonant transitions beginning when doubly-ionized helium recombines and emits a photon from its $n=2$ to $n=1$ electronic transition (He II 4686\,\AA\ arises from the earlier $n=4$ to $n=3$ transition). The ionization step requires a continuum source of photons with energies of at least 54\,eV, suggesting underlying far-UV and X-ray emission even when X-rays are not detected directly \citep{Blagorodnova2017,Leloudas2019}. The TDE-He class, showing only the He II 4686\,\AA\ line, are much rarer, accounting for about 10\% of TDEs. \citet{Hammerstein2023} subsequently identified another rare class, TDE-featureless, with no detectable optical emission lines. Most of the brightest known optical TDEs are in this class. Finally we note that the spectroscopic class of a TDE may depend on the phase at which it was observed: at least one TDE has been seen to evolve from a TDE-H+He to a TDE-He on a timescale of $\sim 100$ days after its peak luminosity \citep{Nicholl2019}.

An obvious question then is what determines the spectroscopic type of a TDE at a given time. Removal of the hydrogen envelope prior to disruption (analogous to a Type Ib supernova), possibly by previous grazing encounters with the black hole, could account for the existence of TDE-He \citep{Gezari2012}, though this cannot explain TDEs that change type. Complete ionization of hydrogen could also explain a lack of Balmer lines \citep{Guillochon2014}. However, the appearance of He II simultaneously with H I in so many TDEs may require fine tuning or appeals to complex geometry, in order to ionize He I (ionization energy 24.6\,eV) without fully destroying H I (13.6\,eV). The most popular explanation for the diversity in line ratios is that a compact photosphere is more optically thick to H$\alpha$ photons (the dominant hydrogen line in the optical) than to He II photons, so that the ratio of escaping photons in each line is determined by the physical size of the reprocessing layer \citep[Figure \ref{fig:roth};][]{Roth2016}. Several statistical studies of TDEs have since shown that the H$\alpha$ luminosity in TDEs is indeed correlated with blackbody radius \citep{vanVelzen2021,Charalampopoulos2022,Hinkle2020}. This correlation seems to confirm the existence of an extended reprocessing layer, but more work is required to understand what determines its size in a given TDE.

\subsection{Spectroscopic and multi-wavelength evidence for outflows and disks}

TDEs also show impressive diversity in the shapes of their emission lines (i.e.~their line profiles). While some are broadly symmetrical in velocity space, others show a blue-shifted peak and a broad red wing. Model line spectra broadened by electron scattering in outflowing gas \citep{Roth2018} give a good match to these profiles, with inferred velocities of $\sim 10^4$\,km\,s$^{-1}$ \citep{Nicholl2020,Charalampopoulos2024}. This is consistent with the wind velocities predicted by \citet{Metzger2016}, and those at low latitudes in the simulations of \citet{Dai2018}. UV spectra have also been obtained for a handful of TDEs. This wavelength range is rich in metal transitions, and again shows evidence for outflows in the form of blue-shifted lines \citep{Cenko2016,Hung2019,Hung2021}. These lines can appear in emission or absorption, depending on their optical depth, and can potentially be used to diagnose the viewing angle through a wind with a polar cavity, as well as the density of the wind \citep{Parkinson2020}.

Many TDEs have also been detected at radio wavelengths, giving a direct measure of the velocity of the fastest material \citep{Alexander2016,Alexander2020,vanVelzen2016a}. Such observations seem to confirm the existence of mass ejected at velocities $\sim 10^4$\,km\,s$^{-1}$. Generally these are launched soon after disruption \cite[e.g.][]{Goodwin2022}, but surprisingly some TDEs appear to launch sudden outflows years after the event \citep{Horesh2021,Cendes2022,Cendes2023}. The combined evidence from the radio, UV and optical observations suggests that outflows are common in TDEs, and natural sites of reprocessing to produce the optical emission. A small fraction of TDEs (only three well-studied events to date) produce much brighter radio and X-ray emission and even gamma-rays, requiring relativistic jets rather than wide-angle winds \citep{Bloom2011,Burrows2011,Zauderer2011,Cenko2012,Pasham2023,Andreoni2022}. Where optical spectroscopy has been obtained, the thermal component in these relativistic TDEs appears similar to the TDE-Featureless class. It should be noted that even among the more common (non-relativistic TDEs), their X-ray and radio light curves are very diverse, and significant work will be required to understand these properties.

Occasionally, TDEs exhibit a very different kind of emission line morphology. An edge-on, rotating disk will emit blue-shifted photons from the approaching side and red-shifted photons from the receding side, leading to a double-peaked line profile. Detections of hydrogen line profiles consistent with both circular \citep{Short2020,Hung2020} and elliptical disks \citep{Wevers2022} shortly after the light curve peaks in two TDEs provide strong evidence that at least some TDEs form their accretion disks quickly. Further evidence for rapid disk formation comes from iron lines, consistent with a hot corona seen in AGN \citep{Wevers2019}. However, it is unclear whether early disk emission is a rare phenomenon, or simply unobservable unless the system has a particularly low density or favorable inclination to allow the direct escape of disk photons \citep{Gomez2020}. Sample studies to date \citep{Charalampopoulos2022,Hammerstein2023} find that TDEs exhibit a range of complex line profiles, and more work is required to follow the time evolution of TDE spectra and model them in detail.

\section{Surprises, outliers and open questions}\label{sec:openquestions}
Here we list a non-exhaustive list of intriguing aspects of TDEs and related transients that go beyond the simpler picture described above, along with relevant citations:

\begin{itemize}
    \item Observations of repeating partial disruptions. These require an initial orbit with lower eccentricity such that the entire star is bound to the black hole after disruption (though still far from circular orbits, $0.9 \lesssim e \lesssim 0.999$) \citep[e.g.][]{Payne2021,Somalwar2023,Wevers2023}.
    \item Observations of related repeating nuclear transients, such as quasi-periodic eruptions (QPEs) in X-rays repeating on timescales of hours-days  
    \citep[e.g.][]{Miniutti2019,Giustini2020,Arcodia2021,Arcodia2024} or weeks \citep{Guolo2024,Evans2023}. These could result from very weak partial disruptions or mass stripping \citep{King2020,Krolik2022,Lu2023}, instabilities in TDE accretion disks \citep{Pan2022,Sniegowska2023,Kaur2023}, or another orbiting body interacting with the disk \citep{Sukova2021,Xian2021,Linial2023,Tagawa2023,Franchini2023}.
    \item Unbound remnants of partial TDEs. Stars that aren't fully disrupted and whose cores are unbound after disruption will leave the black hole at extremely high speeds, potentially becoming observable ``hyper-velocity" stars \citep[e.g.][]{manukian_turbovelocity_2013, Evans2023}. These stars will initially be puffed up from the tidal interaction with the SMBH, and will likely exhibit compositional differences due to the removal of their outer layers \citep[e.g. similar to binary stripped stars,][]{gotberg_spectral_2018}.
    \item Composition anomalies. Some TDEs have been observed to have high abundance ratios of Nitrogen/Carbon, implying the disruption of stars with high levels of CNO cycle material. This could imply a preference for over-massive stars compared to those predicted from two-body relaxation and traditional stellar mass functions \citep[e.g.][]{kochanek_abundance_2016, yang_carbon_2017, mockler_evidence_2022, miller_evidence_2023}. 
    \item TDEs around intermediate mass black holes (IMBHs) and stellar mass black holes. Theoretical expectations for these events differ somewhat from SMBH TDEs. The main reasons for these differences are that most disruptions will be less relativistic (smaller $R_t/R_g$), mass return rates to the black hole will be more super-Eddington ($\dot{M} \propto M_\bullet^{-1/2}$ and $L_{\rm edd} \propto M_\bullet$), and in the case of stellar mass disruptions, the mass of the star will become comparable to the black hole and the approximation that $q << 1$ will no longer be valid. Observational candidates for IMBH and stellar mass TDEs include \citet[][]{angus_fast-rising_2022, nicholl_AT2022aedm_2023}. 
    \item TDEs by SMBH binaries. At sufficiently large separations ($>>$ the sphere of influence), TDE rates and light curves around each individual SMBH should be similar to the rates around single SMBHs. As the binary hardens, the increasingly non-spherical potentials \citep[at separations $\lesssim 50 \times$ sphere of influence, e.g.][]{liu14}, and eventual three-body interactions \citep[at separations $\sim$ sphere of influence, e.g.][]{chen11, Wegg2011, fragione_secular_2018, mockler23} will dramatically increase the TDE rates. At even smaller separations ($\sim$ comparable to orbits of bound debris), the presence of a secondary SMBH can also dramatically influence the light curve evolution, and even lead to accretion onto both black holes \citep{ricarte16}. 
    \item Asymmetries in the gravitational potential. In a non-spherical potential, it generally becomes more efficient to move stars onto disrupting orbits. Torques from the asymmetric portion of the potential can more quickly change the angular momentum of individual stellar orbits than the smaller kicks from two-body relaxation alone, leading to higher TDE rates. 
    This is true for axisymmetric potentials \citep[e.g.][]{vasiliev_loss-cone_2013}, triaxial potentials \citep[e.g.][]{Merritt2004}, and also the more extreme cases of eccentric nuclear disks \citep[e.g.][]{madigan_dynamical_2018} or SMBH binaries (described above).  
    \item Potential connections to AGN. TDE searches have historically excluded AGN in the pursuit of `clean' samples, however there are many reasons to expect that the same processes that lead to increased AGN activity might also lead to increased numbers of TDEs \citep{French2020, dodd_landscape_2021}. Additionally, there has been at least one observation of a TDE candidate in a changing-look AGN (CLAGN), potentially suggesting that TDEs can trigger CLAGN \citep{trakhtenbrot_1es_2019, masterson_evolution_2022}. 
    \item Dust echoes from TDEs, detected in the infrared, appear to be a common feature \citep[e.g.][]{vanVelzen2016,Jiang2021,Newsome2024}. While dusty tori are expected in AGN, the origin of dust in TDEs is less clear. We may also be missing a population of heavily dust-obscured TDEs \citep{Mattila2018,Kool2020}, though TDE searches in the infrared are becoming more systematic \citep{Masterson2024}.
    \item Some of the earliest TDE candidates were detected not through the flare itself, but via its influence on the surrounding environment. The extreme UV and X-ray radiation from a TDE can ionize gas in the host galaxy, leading to emission lines of highly ionized species (e.g. [Fe VII]); these galaxies have been termed Extreme Coronal Line Emitters \citep{Komossa2008,Wang2011,Wang2012}. Now that TDEs are regularly discovered in real time, a few known events have started to display these same lines, providing an opportunity to study the conditions of gas in the nuclei of their host galaxies \citep{Onori2022,Short2023,Clark2024,Hinkle2024}.
\end{itemize}

\section{Summary}

Stars are disrupted by the SMBHs at the centers of galaxies at a rate of $\sim 10^{-5} - 10^{-4}$/galaxy/year. Through observing this process, we learn about the mass function of SMBHs, the limits of black hole growth, and the stars in the nuclei of galaxies. For example, TDE rates are higher in centrally concentrated galaxies with recent star formation, giving us clues as to the properties of the host galaxies on parsec size scales. Most of the stars disrupted in TDEs come from near the sphere of influence of the SMBHs, and are therefore disrupted on very eccentric, nearly unbound orbits. This results in asymptotic mass fallback rates that approach $t^{-5/3}$ power laws for full disruptions and $t^{-9/4}$ power laws for partial disruptions. The fallback rates are also dependent on the properties of the black holes, disrupted stars, and initial stellar orbits. 

The dissipation of the kinetic energy of the gas falling back to the black hole powers bright flares that are often close to the Eddington limit of the SMBH.
The flare timescales and initial decline rates of most TDEs are similar to the mass fallback rates predicted from simulations, though at late times a flattening has been observed in many light curves. This late-time plateau has been modeled as the result of a viscously spreading disk. Theoretically predicted scaling relations with the SMBH mass have been found with the timescale and luminosity of the initial flare, and with the luminosity of the late-time plateau.

Unlike AGN, TDEs exhibit soft, thermal spectra. They are often bright at optical and UV wavelengths, and well-fit by blackbody SEDs that imply emitting areas with radii $\sim 10^{14} - 10^{15}$ cm. While there is still debate over what powers the initial optical and UV flares (shocks? accretion?), the observations of X-rays in some events point to a compact emitting region of order the size of the predicted accretion disk. The spectra of these flares also distinguish TDEs from AGN or supernovae. TDEs exhibit mainly hydrogen, helium, and nitrogen emission lines with broader peaks than most AGN, possibly due to electron scattering in a dense reprocessing layer. Non-thermal line excitations also point to significant UV and X-ray flux not present in most supernovae.

As the number of observations of TDEs grows, one thing remains clear -- surprises abound, and there is still much we don't know. While the focus of this chapter is on what one might call `stereotypical' TDEs, discoveries of surprises and outliers such as repeated TDEs, infrared TDEs, candidate TDEs in AGN or around smaller mass black holes, and potentially related nuclear transients like QPEs are all enriching our knowledge of these transients even as they provide us with new puzzles to solve.

\begin{ack}[Acknowledgments]

The authors would like to think the KITP workshop, ``Towards a Physical Understanding of Tidal Disruption Events" for the opportunity to work on this chapter together in person. This research was supported in part by grant NSF PHY-2309135 to the Kavli Institute for Theoretical Physics (KITP).
MN is supported by the European Research Council (ERC) under the European Union’s Horizon 2020 research and innovation programme (grant agreement No.~948381) and by UK Space Agency Grant No.~ST/Y000692/1.

\end{ack}

\seealso{Space Science Review ``The Tidal Disruption of Stars by Massive Black Holes" \citep[this is a series of 17 in depth review articles, the introduction of which is cited here,][]{jonker_editorial_2021}}

\bibliographystyle{Harvard}
\bibliography{reference}

\end{document}